\documentclass[twocolumn]{article}

\usepackage{geometry}
\usepackage{cite}
\usepackage[utf8]{inputenc}
\usepackage{multirow}
\usepackage{graphicx}
\usepackage{outlines}
\usepackage[dvipsnames]{xcolor}
\usepackage{textcomp}
\usepackage{float}
\usepackage{ulem}
\usepackage{comment}
\usepackage{amssymb}
\usepackage{chngcntr}

\newcommand{\onlinecite}[1]{\hspace{-1 ex} \nocite{#1}\citenum{#1}} 

\usepackage{hyperref}
\hypersetup{
    citecolor = red,
    filecolor = green,
    urlcolor = orange,
    colorlinks = true, 
    linktoc = all,     
    linkcolor = blue,  
}

\title{Considerations for neuromorphic supercomputing in semiconducting and superconducting optoelectronic hardware}
\author{Bryce A. Primavera and Jeffrey M. Shainline}
\date{April 2021}

\begin{document}

\twocolumn[
\begin{@twocolumnfalse}
\maketitle
\begin{abstract}
Any large-scale spiking neuromorphic system striving for complexity at the level of the human brain and beyond will need to be co-optimized for communication and computation. Such reasoning leads to the proposal for optoelectronic neuromorphic platforms that leverage the complementary properties of optics and electronics. Starting from the conjecture that future large-scale neuromorphic systems will utilize integrated photonics and fiber optics for communication in conjunction with analog electronics for computation, we consider two possible paths towards achieving this vision. The first is a semiconductor platform based on analog CMOS circuits and waveguide-integrated photodiodes. The second is a superconducting approach that utilizes Josephson junctions and waveguide-integrated superconducting single-photon detectors. We discuss available devices, assess scaling potential, and provide a list of key metrics and demonstrations for each platform. Both platforms hold potential, but their development will diverge in important respects. Semiconductor systems benefit from a robust fabrication ecosystem and can build on extensive progress made in purely electronic neuromorphic computing but will require III-V light source integration with electronics at an unprecedented scale, further advances in ultra-low capacitance photodiodes, and success from emerging memory technologies. Superconducting systems place near theoretically minimum burdens on light sources (a tremendous boon to one of the most speculative aspects of either platform) and provide new opportunities for integrated, high-endurance synaptic memory. However, superconducting optoelectronic systems will also contend with interfacing low-voltage electronic circuits to semiconductor light sources, the serial biasing of superconducting devices on an unprecedented scale, a less mature fabrication ecosystem, and cryogenic infrastructure. 
\vspace{2em}
\end{abstract}
\end{@twocolumnfalse}
]

\setcounter{tocdepth}{4}
\setcounter{secnumdepth}{4}

\section{\label{sec:introduction}Introduction}
The foundations of cognition remain a great frontier of science, with potentially enormous ramifications for technology and society. A hardware capable of simulating spiking neural networks with the scale and complexity of the brain or even beyond could be a powerful tool in deciphering this enigma. Achieving such large-scale systems has proven to be non-trivial with established CMOS hardware \cite{furber2016large}. A significant challenge will be to enable efficient communication with low-latency amongst billions or trillions of neurons. Optics appears well-matched to the task, as the lack of resistive, capacitive, and inductive parasitics makes optical links more amenable to high fan-out than electrical interconnects \cite{shainline2019superconducting}. While digital systems partially circumvent this issue by leveraging time-multiplexing to artificially increase fan-out \cite{young2019review}, multiplexing introduces latency that scales exponentially above a certain data load \cite{hennessy2011computer}. Optical interconnects may enable direct connections between neurons which would eliminate all traffic-induced delays and support larger, faster, and more interconnected networks. However, while the lack of interaction between photons is beneficial for reducing parasitics during communication, it is a detriment to computation. Electronic circuits are better suited to implement complex, nonlinear neuronal functions. It is reasonable to anticipate performance gains from optoelectronic neural systems leveraging optics for communication and electronics for computation, provided the hardware can be realized.

\begin{figure*}
    \centering
    \includegraphics[scale=1]{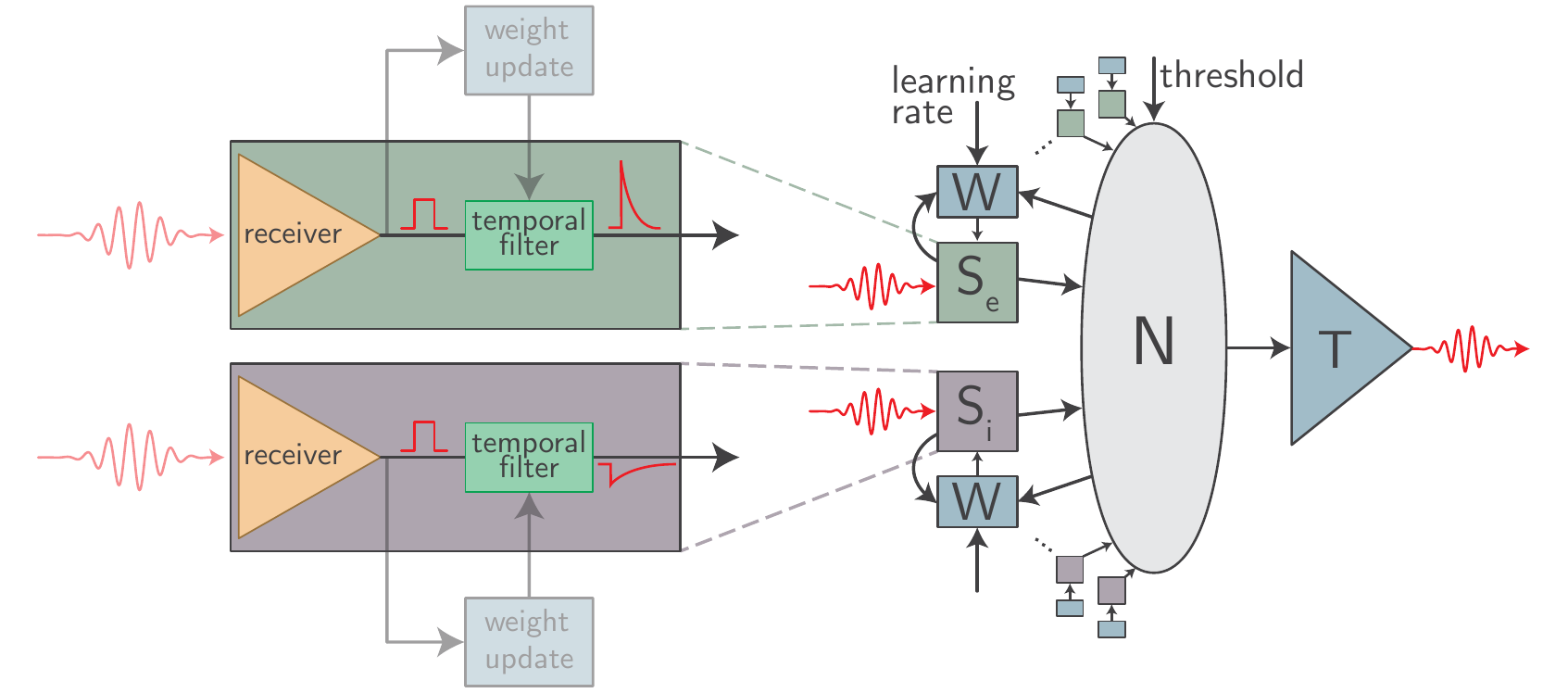}
    \caption{An abstract schematic of the class of optoelectronic neurons meeting our three criteria. Each synapse ($S_e$ and $S_i$ for expiatory and inhibitory synapses respectively) is implemented with a physical circuit block containing a detector and a temporal filter. The detector produces an all-or-nothing electrical pulse upon receipt of an optical spike which is then processed by the filter. The parameters of the filter (time constant, weight, etc.) can be set individually for each synapse. A local weight update circuit (W) implements plasticity mechanisms at each synapse. Synaptic outputs are integrated in the soma (N) which drives an optical transmitter to downstream connections upon reaching threshold.}
    \label{fig:Schematic}
\end{figure*}

Our proposal to fabricate a direct, physical connection between every pair of connected neurons is known as the fully dedicated axon approach to communication \cite{seda2016}. While this strategy requires largely fixing network topology in hardware\textemdash a chief disadvantage when compared with highly reconfigurable digital systems\textemdash the reduced overhead and elimination of communication bottlenecks will greatly benefit performance. We further specify that all synapses, dendrites, and neurons utilize fully dedicated electronic circuits, so that each element of hardware has a one-to-one correspondence with its information-processing role in the neural system. This fully dedicated approach is advantageous if one aspires to create a diverse array of synaptic and dendritic behaviors at each neuron, as observed in biological neural systems \cite{marder1987neurotransmitters, euler2001dendritic}. For instance, a different time constant or plasticity mechanism could be implemented at every synapse on a single neuron. Perhaps more importantly, fully dedicated components eliminate the auxiliary hardware required to perform multiplexing operations. Further, performing synaptic weighting and temporal dynamics in the electronic domain allows for binary optical communication, which minimizes the amount of optical energy per spike and reduces noise incurred by communication. The scope of this paper is therefore limited to networks meeting these three conditions:

\begin{enumerate}
    \item Direct, optical connections are utilized for communication between neurons (fully dedicated axons).
    \item Optical communication is binary. The amplitude of the optical signal carries no information.
    \item All synaptic, dendritic, and somatic computations are performed by fully dedicated electronic circuits.
\end{enumerate}

With these conjectures established, a picture of the hardware under consideration begins to emerge. There is a single optical transmitter at each neuron. This light emitter produces a short pulse of light each time the neuron spikes. The optical pulse is coupled into a waveguide, and optical power is tapped from the waveguide for each downstream synapse. Each synapse contains a photodetector which registers an all-or-nothing synapse event. From there, all synaptic weighting, spike-train filtering, dendritic processing, signal summation, neuronal thresholding, and plasticity mechanisms are implemented in the electronic domain with tailored integrated circuits. A schematic of this general framework is shown in Fig.\,\ref{fig:Schematic}.

There are potentially multiple ways to physically implement this model. The remainder of this paper will discuss two possible implementations\textemdash a superconducting platform and a room-temperature all-semiconductor system. The superconducting platform, known as SOENs (Superconducting OptoElectronic Networks) is discussed in prior work \cite{shainline2017superconducting, shainline2019superconducting, shainline2019fluxonic, sh2021}. In short, optical links are formed from semiconductor light sources and superconducting nanowire single photon detectors (SNSPDs). Computation is performed with analog Josephson junction (JJ) circuits and memory is implemented with persistent current in superconducting loops. The semiconductor implementation is imagined as an exact analogue of the SOENs platform, except without the benefits (or limitations) of cryogenic elements. Traditional photodiodes enable optical communication, analog CMOS circuits provide computation, and emerging memory devices provide synaptic memory.

This paper seeks to analyze the suitability of both platforms for implementing large-scale optoelectronic neuromorphic networks. Despite limiting our discussion only to architectures meeting our three conjectures, there remains a vast space of design choices, making it difficult to draw hard-and-fast conclusions. Nevertheless, interesting guidelines can be obtained by analyzing limits of technologies most likely to be used in each platform. Important benchmarks for device performance are also identified, which may be of use in monitoring the development of this field.

\section{\label{sec:communication}Communication}
\subsection{Optical Receivers}
We begin analysis of optical interconnects with receivers. There are two ways the receiver influences the power budget of an optical link: (1) The receiver (and the electrical components it must drive) sets the minimum optical signal that must be produced by the light source, and (2) the receiver may require electrical power of its own to run. It is found that the energy per spike may be quite similar in both platforms once cooling is accounted for in the superconducting case. However, the optical power required from light sources is reduced by a factor of 1000 in the superconducting case, at least when compared to the semiconductor receivers of comparable total efficiency, which omit transimpedance amplifiers \cite{miller2017attojoule}.

\subsubsection{Superconductor Receivers}
\begin{figure}[!h]
    \centering
    \includegraphics[width=.49\textwidth]{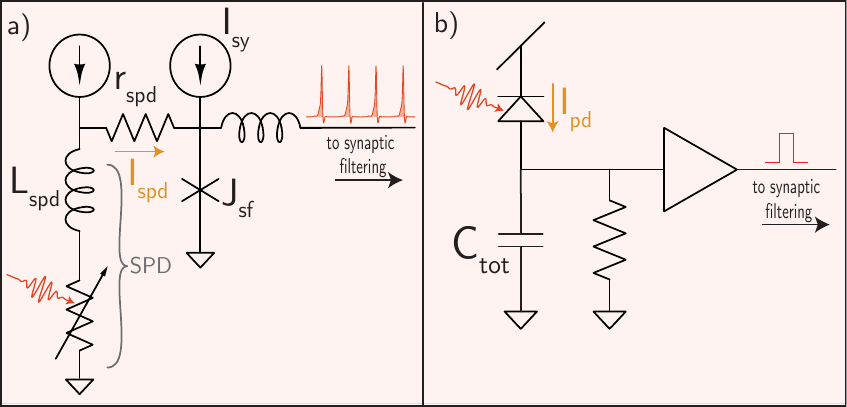}
    \caption{Receivers for the (a) superconducting and (b) semiconducting platforms. Note that synaptic weighting for the semiconductor case is included in the filtering circuitry, shown in figure\,\ref{fig:filtering}(b).}
    \label{fig:sup_synapse}
\end{figure}

As stated previously, the SOENs platform utilizes SNSPDs to detect optical signals as faint as a single photon. Physically, an SNSPD is a superconducting nanowire biased with a current source ($I_{\mathrm{spd}} \approx$ 10\,\textmu A). The simple structure makes fabrication and waveguide integration straightforward \cite{spga2011,pesc2012,akhlaghi2015waveguide,feka2015,saga2015,shbu2017b,ferrari2018waveguide,buta2020}. Photons travelling through a waveguide evanescently couple to a nanowire on the surface of the waveguide. A single photon has enough energy to drive the nanowire from the superconducting phase to a resistive state. In SOENs receivers, this momentarily redirects the bias current along an alternate conduction pathway that activates a JJ circuit to register the synapse event and conduct further synaptic processing (Fig.\,\ref{fig:sup_synapse}(a)).

While an SNSPD itself dissipates zero static power, electrical power is still required for superconducting receivers. Current biases will require some power, but should be shared by many devices (Sec.\,\ref{sec:soma}), ameliorating the cost. More important is dynamic electrical power consumption associated with detection events. The nanowire has an inductance, $L_{\mathrm{spd}}$, that stores energy from the current bias. During a detection event, this energy is dissipated in the resistor $r_{\mathrm{spd}}$. The electrical energy necessary to detect each photon is then $\frac{1}{2}L_{\mathrm{spd}}I_{\mathrm{spd}}^2$. $L_{\mathrm{spd}}$ can be as low as 100\,nH, resulting in an electrical energy consumption ($E_{\mathrm{spd}}$) of around 5\,aJ/spike.

Since an SNSPD is capable of detecting single photons, it will operate near the quantum limit of optical communication \cite{razavi2012design}. We assume that the detection of a single photon will be treated as the registering of a synaptic event. The probability of a light source producing a spike with a certain number of photons within a fixed time window is given by a Poisson distribution. We will also conservatively assume a detection efficiency $\eta_D$ of 70\% (higher detection efficiency is certainly possible \cite{mave2013,rene2020}). The probability of measuring zero photons during a spiking event is then given by:
\begin{equation}
    P(0) = \sum_{k=0}^{\infty} \frac{N_{ph}^k e^{-N_{ph}}}{k!}(1-\eta_D)^{k} = e^{-N_{ph}\eta_D},
\label{eq: poisson}
\end{equation}
where $N_{ph}$ is the average number of photons per spiking event. Neural systems are known for remarkable robustness to and even utilization of noise \cite{stein2005neuronal, mcdonnell2011benefits}. Detecting only 99\% of spikes may be tolerable and would still represent a significant improvement over biology, wherein synapse reliability is typically in the range of 5\% - 80\%  \cite{allen1994evaluation,li1997}. From Eq.\,\ref{eq: poisson}, this would correspond to roughly 7 photons (0.9\,aJ for $\lambda = 1.5$\,\textmu m) needed to reach the receiver. The total number of photons produced by the source will need to be higher to account for energy losses in the link. The total optical energy per spike, $E_{\mathrm{opt}}$, will be:
\begin{equation}
    E_{\mathrm{opt}} = \frac{N_{\mathrm{ph}} h \nu}{\eta}.
\end{equation}
$h\nu$ is the energy of a single photon and $\eta$ is the total energy efficiency of the optical link. $\eta$ includes all optical losses and the inefficiency of the transmitter. This efficiency factor will be highly dependent on the specifics of the platform, but for now we will leave it as a free variable. The total power consumed by the optical link is the sum of $E_{\mathrm{opt}}$ and $E_{\mathrm{spd}}$. Accepting a 1\% error rate, these two contributions to the total energy will be roughly equal when $\eta = 20\%$. Such a high efficiency is likely near the limits of physical possibility. For more realistic values of $\eta$, $E_{\mathrm{opt}}$ will dominate.

Importantly, superconducting electronics come with a cooling overhead (Sec.\,\ref{sec:instantiation}). We conservatively assume that every watt of power produced at low temperature will require 1\,kW of refrigeration power. System-level effective optical energy per spike for superconducting links will be no less than 1\,fJ.

Fabrication of waveguide-integrated SNSPDs has become commonplace in recent years \cite{spga2011,pesc2012,akhlaghi2015waveguide,feka2015,saga2015,shbu2017b,ferrari2018waveguide,buta2020}. SNSPD materials include NbN, NbTiN, WSi, and MoSi. Superconducting films (3\,nm - 10\,nm) can be sputtered at room temperature atop many substrates and patterned into wires from 50\,nm to 5\,\textmu m wide using conventional lithography and etching. Multiple planes of SNSPDs have also been demonstrated \cite{vema2012}\textemdash a promising development for future large-scale neuromorphic systems (Sec.\,\ref{sec:instantiation}). Waveguide-integrated NbN SNSPDs can reach photon count rates exceeding 1\,GHz \cite{rosenberg2013high,vetter2016cavity}. However, slower detectors, such as MoSi and WSi SNSPDs with 20\,MHz count rates, have demonstrated the best yields to date (99.7\% \cite{wove2019}). Previous statements that SOENs were limited to 20\,MHz were motivated by these pragmatic concerns about the current state of fabrication \cite{shainline2019superconducting}.

\subsubsection{Semiconductor Receivers}
While semiconductor receivers are the predominant technology for long-distance optical communication, intra-chip optical receivers deviate significantly from their long-distance counterparts, as traditional transimpedance amplifiers likely consume too much electrical power, despite impressive optical sensitivities. This has led to the proposal of ``receiverless'' designs that omit amplifiers altogether \cite{miller2017attojoule}. Receiverless communication uses a photodetector to directly drive the input of CMOS gates. Photons produce electron-hole pairs in the photodetector, which in turn charge the CMOS gate capcitance up to the switching voltage. A circuit diagram of the scheme is shown in Fig.\,\ref{fig:sup_synapse}(b) in which a photodiode directly drives a CMOS digital buffer. A resistor is also placed in parallel to allow the receiver to reset. In principle the resistor is unnecessary if an optical reset is used as described in \cite{debaes2003receiver}. The resistor would increase the minimum optical power necessary to register a spike and limit the bandwidth of the receiver. 

With optical link efficiency $\eta$, the necessary optical energy required to drive the receiver to a voltage $V$ is \cite{miller2017attojoule}:
\begin{equation}
\label{eq: semi}
    E_{\mathrm{opt}} = \frac{C_{\mathrm{tot}} V}{\eta \mathcal{R}}.
\end{equation}
$\mathcal{R}$ is the responsivity of the detector, typically of order 1\,A/W. $C_{\mathrm{tot}}$ includes the photodiode capacitance, the CMOS gate capacitance, and any wiring capacitance. It is reasonable to consider values for $C_{\mathrm{tot}}$ at the femtofarad level. For 1.5\,\textmu m photons and a required voltage swing of 0.8\,V, $E_{\mathrm{opt}} \approx 0.7 $\,fJ (5000 photons) for unit efficiency. This is similar to the superconducting case, once cooling is considered. If two optical communications links were identical in all measures (source efficiency, optical losses, etc.) except one was cooled to 4\,K with SNSPDs and the other operated at room-temperature with photodiodes, then communicating a spike would cost nearly the same energy at the system level in each link. The power required for cryogenic cooling pays for itself with reduced light levels in the optical link. Cooling semiconductor receivers to 4\,K does not appreciably improve the situation, as the number of photons required in the receiverless case is related to charge, capacitance, and voltage, not thermal noise. For capacitances below 1\,fF (a difficult task), semiconductor receivers could potentially consume even less energy than their superconducting counterparts. Waveguide-integrated femtofarad photodiodes have been demonstrated in both SiGe and Ge \cite{derose2011ultra}. Polysilicon photodiodes are also attractive for increased manufacturability \cite{meor2014}. Most photodiodes have far better speed than required for neuromorphic applications, reaching up to 45\,GHz \cite{derose2011ultra}.

Just as with SNSPDs, semiconductor receivers will also require electrical power, even if it is minimized by the receiverless approach. In this case, there will be static power dissipation through the leakage current of the photodiode. Assuming a 1\,V bias, a leakage current on the order of 1\,nA \cite{zhang2020scalable}, and an optical link efficiency of 1\%, this static dissipation would dominate power consumption for average spiking rates below 10\,kHz. The development of low capacitance, zero-bias photodiodes \cite{nozaki2018forward} would be a major advantage towards making efficient, low frequency networks. Static power consumption is also a major question for many avalanche photodiode (APD) receivers. Avalanche gain could provide a significant (at least one order of magnitude) reduction in the necessary optical power per spike \cite{miller2017attojoule}. While often associated with higher bias voltages, germanium waveguide-integrated avalanche detectors have been demonstrated to provide 10\,dB of gain even at 1.5\,V bias \cite{assefa2010reinventing}. However, dark current is still typically in the microamp range for such detectors \cite{assefa2010reinventing,virot2014germanium}, meaning that brain-scale networks are likely out of reach due to power constraints (Sec.\,\ref{sec:instantiation}). APDs may be of interest in smaller, faster spiking networks, however. Another intriguing possibility is to reduce static power consumption through cooling, as the dark current could potentially be reduced by orders of magnitude \cite{pizzone2020analysis}. However, in that case one forfeits a major advantage of the semiconductor approach.

\begin{figure}[h!]
    \centering
    \includegraphics[scale=1]{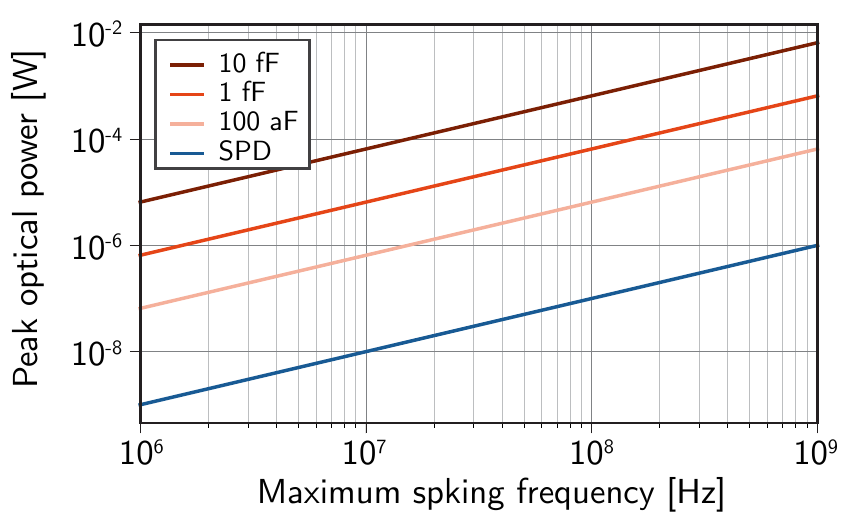}
    \caption{The required optical power to drive 1000 downstream synapses within one inter-spike interval for a given spiking frequency assuming receiverless photodiodes with optical link efficiency $\eta = 1$.}
    \label{fig:communication}
\end{figure}

While the receiverless scheme is promising for achieving low energies per spike, it places significant burden on the transmitter side of the link. Neuromorphic applications magnify this burden, as neurons are expected to drive thousands of downstream connections in parallel. Additionally, the receiver capacitance must be charged quickly to maintain high spiking frequencies. The result is that relatively large optical power is required from transmitters. The best case ($\eta = 1$) scenario is shown in figure \ref{fig:communication}. Semiconductor receivers can be expected to require around one thousand times the optical power of superconducting receivers and the highest spiking frequency of a neuron could very well be limited by the power output of the light source. The ramifications of this result on prospective light sources are discussed in the next section. 

\subsection{Optical Transmitters}
The transmitter is expected to dominate the power budget of optical links for both platforms. Room-temperature, CMOS-integrated light sources have been a holy grail for decades, but materials integration issues have kept this prized objective out of reach. For superconducting systems, SNSPDs drastically lower the power requirements of light sources, while cryogenic temperatures improve light source efficiency. Light sources are likely significantly simpler in the superconducting case. However, interfacing low-voltage superconducting electronics with semiconductor light sources \cite{mccaughan2019superconducting} presents an obstacle that is absent from the all-semiconductor platform.

\subsubsection{Integrated Light Sources}
Optical coherence is not a requirement for the envisioned system. NanoLEDs are thus an attractive option due to their ease of fabrication, lack of threshold current, and improving efficiency with shrinking scale \cite{romeira2019physical}. However, nanoLEDs struggle to produce optical power significantly greater than 1\,\textmu W \cite{romeira2019physical}. While semiconductor systems targeting spiking frequencies in excess of 1\,MHz may be forced to turn to lasing, nanoLEDs should be more than sufficient for superconducting platforms. Either way, integrating millions of light sources on a 300\,mm wafer remains highly challenging. The indirect band gap of silicon drastically reduces light emission. Off-chip light sources are used in some applications, but are likely untenable for massive systems, as their high static power consumption is incommensurate with the sparsity of neural activity. Integrated light sources would be a tremendous boon, if not a requirement for the success of large-scale optoelectronic neuromorphic computing. There are two courses of action: (1) force silicon to emit light through either material and/or environmental modifications or (2) integrate direct bandgap materials on silicon. 

Many strategies towards silicon light sources have been pursued \cite{iyxi1993,shxu2007} including quantum confinement in Si-based superlattices \cite{warga2008electroluminescence} and nanocrystals \cite{wabo2005}, emission from embedded erbium \cite{enpo1985,paga1996}, point-defect emitters \cite{brha1986,brbr1989,rosh2007b,bata2007}, extended defects \cite{ng2001efficient}, strain dislocations \cite{kvba2004}, and engineering of the local density of optical states \cite{grzh2001}. Total efficiency from 0.1\% \cite{kvba2004} to 1\% \cite{grzh2001} has been demonstrated at room temperature, but not at powers and areas suitable for the semiconductor receivers introduced in the previous section.

Abandoning silicon as an active optical element, many researchers turned towards epitaxial germanium grown on Si \cite{sun2009toward}. Like silicon, germanium is an indirect-gap semiconductor. However, the direct gap is only 136\,meV higher than the indirect than the indirect gap, and clever implementation of strain \cite{ishikawa2003strain,ghrib2012control,tani2021enhanced} and heavy $n$-type doping \cite{liu2007tensile,el2009enhanced,sun2009direct,camacho2013direct,virgilio2013radiative} can lead to appreciable direct, radiative recombination. These efforts have led to Ge-on-Si lasers \cite{sun2009room,liu2010ge}, but it has proven difficult to reduce the threshold current and increase device efficiency. Another approach is to grow SiGe with a hexagonal lattice on GaAs, leading to a direct gap \cite{Fadaly2020}, but this does little to solve integration problems.

At present, neither Si nor Ge emission has proved satisfactory for the needs of digital communication, so integrating III-V materials on silicon substrates has received significant attention. Pending a watershed moment in silicon sources, III-V integration will be required for the semiconductor platform (although not necessarily in the superconductor case, where low-temperature changes the physical context). Epitaxial growth would be an attractive solution for III-V integration due to the high throughput \cite{norman2018perspective}, but defects due to lattice mismatch have so far prevented this method from large-scale adoption. III-V quantum dots are more robust to such defects and have demonstrated high optical powers with small footprints \cite{chli2016,jung2017high, norman2018perspective}, albeit typically grown on offcut Si substrates that are not CMOS compatible or with thick buffer layers that make optoelectronic contact difficult. More work is required to realize scalable, cost-effective integration of III-V quantum dot light sources with CMOS electronics, passive photonic waveguides, and efficient photodetectors. Without epitaxial growth, the semiconductor platform would be less scalable due to the limited size of III-V wafers and the expense of performing wafer bonding. A variety of schemes have been proposed \cite{norman2018perspective,tang2019integration}, including die-level bonding \cite{sost2016,crsa2017}, wafer-level bonding \cite{huli2019,szha2019,jito2020}, transfer printing \cite{jubo2012,zhha2018,zhang2019iii}, and selective-area epitaxy \cite{haxu2021}, but these approaches still appear cumbersome when seeking the scale of integration considered here.

The situation is significantly more favorable for cryogenic systems. Low temperature often reduces non-radiative recombination \cite{gurioli1991temperature, dolores2017waveguide, sa1958}, improving efficiency for both silicon and III-V light sources. The case of Ge at low temperature is more subtle due to the pecularities of the pseudo-direct gap and inter-valley scattering that is more prevalent at higher temperatures \cite{sun2009toward}. The benefits are further compounded by the low optical power requirements of SNSPDs. When integrating III-V light sources with CMOS, the light sources must be integrated on top of the electronics after the high-temperature dopant activation steps have been performed. Superconductor electronics have no such high-temperature processing steps, so the light sources can be produced on a Si wafer before the electronics are realized. Problems related to offcut Si wafers and thick buffer layers are eliminated. Additionally, silicon light sources, with their superior potential for integration, demand exploration with the superconducting platform. Several silicon point defects typically quenched at room-temperature emerge as narrow-linewidth candidates for light sources in the telecommunications band \cite{davies1989optical,suku2014,buckley2017all,bere2018,chbe2018}. While single-photon emission \cite{hobe2020,redu2020,bech2020} is not the objective in the present context, the narrow linewidth is also attractive for further efficiency gains via the Purcell Effect \cite{romeira2018purcell}. LEDs have already been demonstrated with the W-center defect \cite{buckley2017all,bao2007point}, albeit with poor ($10^{-6}$) efficiencies, limited by electrical injection efficiency rather than emitter lifetime. Photoluminescence studies are promising for orders of magnitude improvement \cite{buckley2020optimization}, but more work is required to improve emission efficiency in an integrated-circuit context. If cryogenic silicon light sources become viable, the superconducting platform might hold a major scalability advantage over the semiconducting analogue. 

\subsubsection{Driving Circuitry}
Both platforms require neurons to drive semiconductor light sources. The transmitter circuitry is thereby required to produce voltages on the scale of the bandgap of the optical source ($\approx 1$\,V). CMOS circuitry, itself a semiconducting technology, naturally operates on this voltage, rendering the driving circuitry a non-issue. Standard MOSFET LED or modulator driving circuits \cite{bowers2016recent, halbritter2014high} can be straightforwardly adapted for neuromorphic applications. Superconductors, however, operate in an entirely different regime, with signals usually on the order of the superconducting energy gap ($\approx 1$\,mV). The optimal method for interfacing superconducting electronics with semiconductor devices is still an area of active research. Recent progress has been made with devices utilizing the massive change in impedance during a phase transition between superconducting and resistive states. In Ref.\,\onlinecite{mccaughan2019superconducting}, a resistive element was heated using 50\,mV pulses to thermally trigger a transition in a superconducting meander. The meander transitioned to a state with resistance in excess of 10\,M$\Omega$ and was used to drive a cryogenic silicon light source waveguide-coupled to an SNSPD. While these results are promising, the light source was only pulsed at 10\,kHz (due to poor source efficiency) and was fabricated on a separate chip. More work is needed to improve the speed, efficiency, and to monolithically integrate driving circuitry with LEDs.

\section{\label{sec:soma}Electronic Neuronal Computation}
Electronic circuitry capable of performing neuronal dynamical operations will also be necessary. Biological neurons are increasingly recognized as sophisticated computational units \cite{kose2000,stsp2015,haah2016,sava2017}. Emulating such complicated behavior has been the subject of extensive investigation in both semiconducting \cite{voma2007,indiveri2011neuromorphic,pfgr2013,brne2013,bega2014,abta2019} and superconducting platforms \cite{shainline2019fluxonic, crotty2010josephson, toomey2019design}. We do not attempt a comprehensive review of circuitry, but rather draw attention to issues specific to optoelectronic networks in both cases.

\subsection{Semiconductor Electronics}
The maturity of CMOS processing has allowed great strides in neuromorphic computing. While optical communication would likely also be advantageous in digital approaches, we focus on analog CMOS neurons for their perceived efficiency advantages \cite{mead1990neuromorphic,rajendran2012specifications}. At a basic level, a neuron must perform three mathematical functions: summation of synaptic inputs, temporal filtering, and threshold detection leading to action potential generation. Summation can be achieved by exploiting Kirchoff's current law. Filtering can be implemented with elementary resistor-capacitor circuits. Thresholding is a natural function of transistors. Building upon this basic mapping, analog neurons have demonstrated a litany of biologically-inspired models \cite{indiveri2011neuromorphic,lide2015}.

It was found in the previous section that optical communication requires a minimum of about 1\,fJ of energy to deliver a spike signal to each synapse. For realistic optical link efficiencies, this value will be at least an order of magnitude larger. Synaptic processing circuits would therefore ideally operate with an energy budget of 10\,fJ - 100\,fJ to process a single spike. Somatic computation could comfortably consume power larger than that of synaptic processing by a factor of the average fan-out (perhaps 1000). Many low-energy neuromorphic demonstrations are promising for reaching these targets. By reducing the membrane capacitance and supply voltage, a neuron capable of 25\,kHz spike rates was demonstrated to consume only 4\,fJ/spike \cite{sourikopoulos20174}. Many other analog neurons, with energies ranging from femtoJoules to picoJoules per spike, fall comfortably below the power consumption of optical communication \cite{indiveri2019importance}. However, it remains to be seen if more complicated neurons and synapses, implementing a critical subset of behavior necessary for cognition, will be able to maintain such low power operation. In terms of speed, CMOS neurons have demonstrated spike rates in excess of 100\,MHz \cite{schemmel2017accelerated}.  Optical communication should face few issues achieving such speeds, \textit{if} sufficiently bright light sources can be efficiently integrated with CMOS circuits.

One challenge for the CMOS approach has been to design compact circuits with long time constants. Long time constants are important for systems targeting biological time scales (upwards of 500\,ms) \cite{indiveri2019importance} or power-law distributions of timescales to implement critical behavior \cite{be2007}. Subthreshold transistor circuits operating with currents in the femtoamp to picoamp range minimize the size of capacitor needed to implement a specific time constant \cite{indiveri2011neuromorphic}. The area constraints of this scheme are discussed in Appendix \ref{apx:TimeCon} and compared to the superconducting approach. 

For a concrete example, a circuit diagram for a memristor implementation of the popular differential-pair integrator (DPI) synapse is shown in Fig. \ref{fig:filtering}(b) \cite{dalgaty2019hybrid}. The DPI produces a decaying exponential post synaptic signal in response to an input voltage pulse\textemdash potentially from an optical receiver. This leaky integrator behavior is characterized by a time constant set by the value of the filtering capacitance and the rate of leakage off the capacitor \cite{chicca2014neuromorphic}. The time constant could potentially be programmed using memristors\textemdash an advantage over superconducting circuits that have been proposed to date.

\subsection{Superconducting Electronics}
Superconducting neurons have been studied nearly as long as CMOS implementations, with a mapping between neuronal functions and superconducting electronics identified in the early 1990s \cite{hago1991, hiak1991}. In this case, Faraday's Law, governing the addition of magnetic flux through mutual inductors to superconducting loops provides the necessary synaptic summation function. Filtering is achieved through resistor-inductor blocks (or RC circuits in some cases \cite{crotty2010josephson}). Josephson junctions (JJs) provide the requisite nonlinear thresholding element.

Like their CMOS counterparts, many superconducting circuits have now been designed to implement sophisticated neuronal dynamics. Superconducting neuromorphic circuits have been designed to implement a variety of bio-inspired neuron models \cite{crotty2010josephson, toomey2019design, schneider2018tutorial}, dendritic processing \cite{shainline2019fluxonic}, and have performed image classification in simulation \cite{schneider2017energy}. The natural spiking behavior of JJs may even require a lower device count than analogous CMOS circuits for various leaky-integrate-and-fire models \cite{crotty2010josephson}. In short, it does not appear that superconducting circuits are any less capable of complex neuronal computation than CMOS, although experimental demonstrations lag far behind.

Superconducting electronics has long been pursued for gains in energy efficiency and speed. Indeed, superconducting elements dissipate zero static power and spike energies are frequently reported in the sub-femtojoule range, including refrigeration. Optical communication is likely to dominate power consumption for superconducting optoelectronic systems (Appendix \ref{apx:squid_area}). In terms of speed, fully electronic superconducting neurons may be capable of spike rates up to 100\,GHz \cite{schneider2018tutorial, schneider2017energy}. However, this is orders of magnitude faster than any SNSPD can respond. This speed disparity is a notable difference between the superconducting and semiconducting architectures. While optical communication could be integrated with CMOS neurons with no degradation in speed, optoelectronic superconducting systems will likely be significantly slower than their fully electronic counterparts. This may be the cost of highly connected systems. That said, the extraordinary switching speed of JJs is still leveraged in optoelectronic networks to perform analog computations within synapses, dendrites, and neurons.

\begin{figure}[h!]
    \centering
    \includegraphics[width=.48\textwidth]{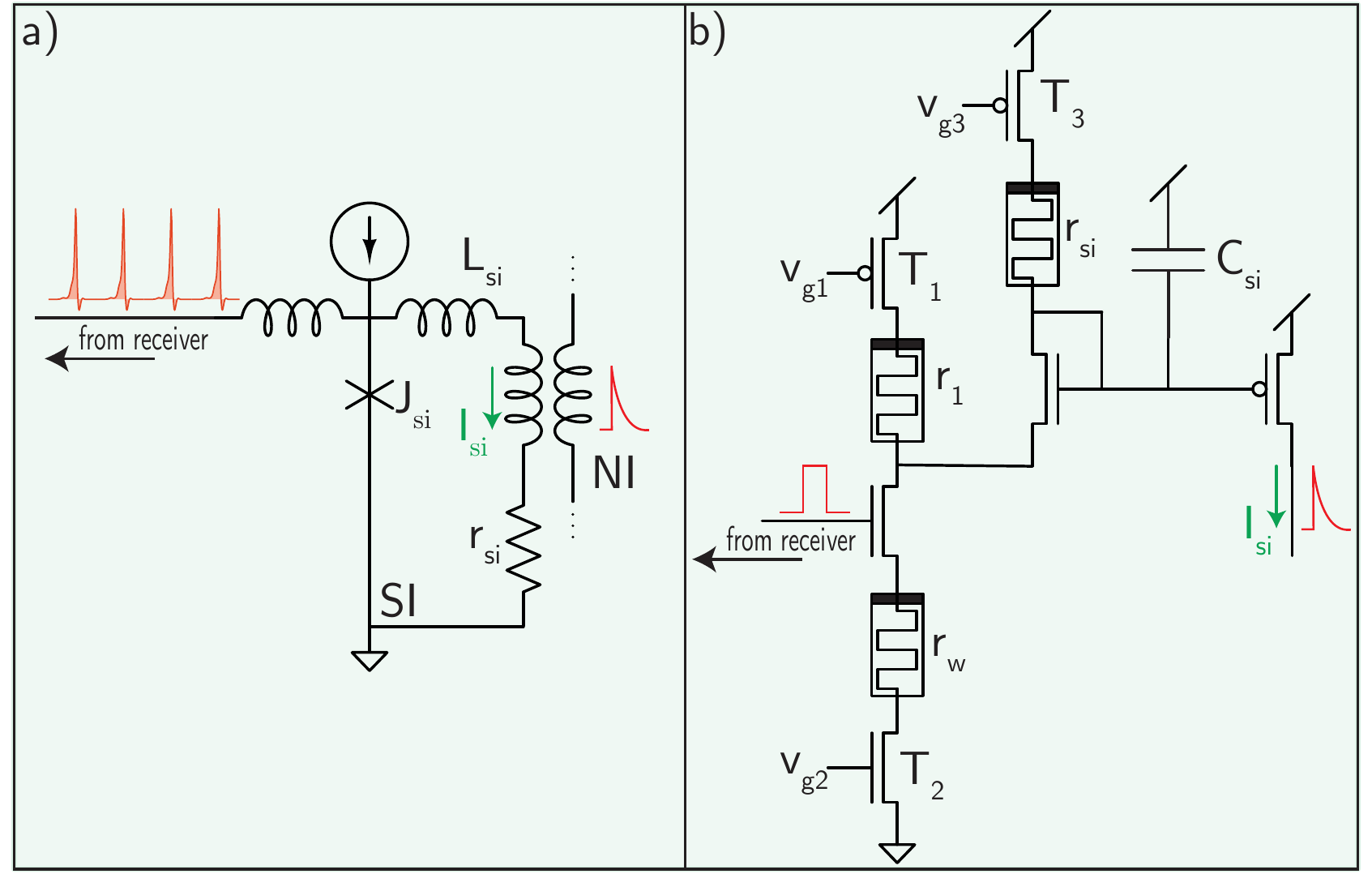}
    \caption{Synaptic filtering circuits for the superconductor (a) and semiconductor (b) cases. Weighting in the superconducting case was shown in Fig.\,\ref{fig:sup_synapse}. The memristor-integrated DPI circuit pictured here is introduced in Ref.\,\onlinecite{dalgaty2019hybrid}. }
    \label{fig:filtering}
\end{figure}

The ability of superconducting electronics to go slow might be just as compelling as their ability to go fast. While it can be challenging to implement long, biologically realistic time constants in CMOS neurons, superconducting loops can create time constants orders of magnitude higher than biology by adjusting the $L/R$ ratio in synaptic and neuronal loops (See Fig.\,\ref{fig:filtering}(a) and Appendix \ref{apx:TimeCon}). The ability to generate dynamics across many orders of magnitude in time also dovetails nicely with suggestions that critical behavior is important for cognition \cite{cocchi2017criticality}. 

Fan-in has traditionally been considered a liability of superconducting electronics. If this were the case, it would clearly be an impediment to mature superconducting neuromorphic systems. For superconducting neurons designed to use single fluxons as synaptic signals, fan-in has recently been analyzed \cite{schneider2020fan}, and it has been found that if a single synapse must be able to drive a neuron above threshold, fan-in may be limited to around 100. However, it is often not necessary for each synapse to be able to trigger a neuronal spike event. It has been analyzed elsewhere that if analog signals containing multitudes of fluxons are communicated from synapses to the neuron cell body, fan-in can likely scale to biological levels through the use of mutual inductors \cite{shainline2019superconducting}. Using more fluxons comes with larger power consumption, but for optoelectronic systems, light production will likely still dominate.

While most diagrams of superconducting circuits (including those here) show many separate biases delivering current to various elements, the ability to construct circuits that can be biased in series will be critical to the scalability of this hardware. A separate bias for every synapse would be untenable in large-scale systems \cite{tolpygo2016superconductor}. This mimics the evolution that occurred in superconducting digital electronics, in which the field has turned away from parallel biasing schemes and embraced serially biased platforms \cite{tolpygo2016superconductor} and current recycling schemes \cite{kisa2011}. SOENs are potenially amenable to serial biasing, but this important point demands further analysis.

A superconducting synaptic filtering circuit is shown in Fig.\,\ref{fig:filtering}(a). Synaptic weighting is implemented in the receiver circuit (Fig.\,\ref{fig:sup_synapse}(a)), so this circuit block is only responsible for converting a train of fluxons into a decaying exponential post-synaptic potential reminiscent of biological and CMOS synapses. A resistor, $r_{\mathrm{si}}$, converts a superconducting persistent current loop into a leaky-integrator in a similar manner to the DPI synapse. The time constant is set by $L_{\mathrm{si}}/r_{\mathrm{si}}$, and the synaptic current can be added to a neuronal circuit through mutual inductors. Unlike the DPI synapse, this circuit does not have a programmable time constant, but does hold the potential to implement a wide range of different time constants by fabricating different values of $L_{\mathrm{si}}$ and $r_{\mathrm{si}}$.

\section{\label{sec:memory}Synaptic Memory}
It has been apparent to the neuromorphic community for some time that large-scale neural systems will require innovative approaches to synaptic memory. A local, analog memory element unique to every synapse will provide the most efficient performance by eliminating memory retrieval and digital conversion. Important metrics for analog synaptic memory technologies include weight precision, volatility, area, write energy, write speed, and endurance (the effective number of cycles in a device's lifetime). We attempt to provide desired benchmarks for a few of these metrics in the specific case of optoelectronic networks. For this section, we assume a speedup of about $10^4$ over biology, for an average spike rate of 10\,kHz and a maximum of 10\,MHz. This is commensurate with both the maximum count rates of high-yield SNSPDs and some of the fastest CMOS electronic neuromorphic systems built to-date.

\subsection{Memory Benchmarks}

\subsubsection{Endurance}
Large-scale neural systems require significant investments in money and time. Operational lifetimes on the scale of decades ($10^9$ seconds), if not longer, are therefore essential. Such systems will be expected to learn continually during that lifespan, placing significant requirements on the durability of memory technologies. The number of times a synapse is updated in its lifetime is a function of neuron spiking frequency ($f$) and the number of synapses that are typically updated after each post-synaptic spike. Neuroscientific evidence has been presented that the number of active presynaptic inputs required to trigger a postsynaptic spike goes as $\sqrt{N}$, where $N$ is the fan-in of the neuron\textemdash exceeding 1,000 for brain-like systems \cite{vrso1996,vora2005}. We assume all synapses that contributed to the spiking of the post-synaptic neuron are updated with each spike. We then estimate the number of weight updates ($N_{\mathrm{update}}$) in the synapses's lifetime ($L$) will be:
\begin{equation}
    N_{\mathrm{update}} = \frac{Lf}{\sqrt{N}}
\end{equation}
For a decades-long lifetime, and a mean spiking frequency of 10\,kHz, the total number of weight updates will be $10^{11}$. This is a challenging demand for many emerging non-volatile memory technologies.

\subsubsection{Update Energy}
One would like the power dedicated to weight updates not to exceed the power used for optical communication. Once again invoking the assumption that $\sqrt{N}$ synapses are updated with each postsynaptic spike, we arrive at the following relation between the energy to produce a single spike ($E_{\mathrm{opt}}$) and that to update a single weight ($E_{\mathrm{update}}$):
\begin{equation}
    E_{\mathrm{update}} < \sqrt{N}E_{\mathrm{opt}}
\end{equation}
Using the analysis in Sec.\,\ref{sec:communication}, 1\,fJ of energy needs to be delivered to the receiver in either platform. Assuming a transmitter efficiency of 1\%, this would mean $E_{\mathrm{opt}}$ is 100\,fJ. Therefore, for a fan-in of 1,000 synapses, $E_{\mathrm{update}}$ would ideally be no more than about 3\,pJ. This value includes any energy consumption of peripheral circuitry, both static and that associated with programming. This efficiency appears to have already been met by several emerging memory technologies \cite{zahoor2020resistive,schneider2018ultralow}.

\subsubsection{Update Speed}
An ideal system would be capable of implementing synaptic updates within the minimum inter-spike interval. While semiconductor optoelectronic systems could potentially produce spike rates in excess of 10\,GHz (assuming sufficiently bright, integrated light sources can be achieved), synapses might need to be taken offline during WRITE operations, as it is unlikely that sophisticated plasticity mechanisms can be implemented in under 100\,ps. Lower maximum frequencies would allow plasticity to be implemented without ever neglecting a spiking event. For our 10\,MHz target, we desire memory updates in under 100\,ns. Slower updates may not be completely intolerable, if network dynamics are robust to missed spikes during synaptic updates or to synaptic weights that are in the process of being altered.

\subsubsection{Weight Precision}
The necessary weight precision will be determined by the specifics of a chosen learning model and the desired application. Weight precision has been the subject of much discussion. It has been suggested that 4-bit precision is sufficient for state-of-the-art mixed signal neuromorphic systems \cite{pfeil20124}. Deep learning systems have also demonstrated success with 8-bit precision\textemdash a significant reduction from 32-bit floating point numbers \cite{wang2018training}. Hippocampal synapses in rats have been inferred to allow at least 26 different states ($\approx 5$ bit), which squares nicely with computer science findings \cite{bartol2015nanoconnectomic}. It has also been argued that metaplasticity mechanisms are more important for lifelong learning than the bit-depth of the synapse \cite{fudr2005,fuab2007}.

Target values for these key synaptic memory metrics are summarized in Table \ref{tab:memory_metrics}.

\begin{table}[h!]
  \begin{center}
    \renewcommand{\arraystretch}{1.5}
    \begin{tabular}{l|c} 
      \textbf{Metric} & \textbf{Goal} \\
      \hline
      Endurance & $>10^{11}$ updates \\
      Update Energy & $<$ 3 pJ\\
      Update Speed & $<100$ ns \\
      Weight Precision & 4-8 bits
      
    \end{tabular}
    \caption{List of desired performance metrics for synaptic memory in a system with average fan-out of 1000, maximum spike rate of 10 MHz, average spike rate of 10 kHz, and spike energy of 100 fJ.}
    \label{tab:memory_metrics}

  \end{center}
\end{table}

\subsubsection{Programming Signals}
One important criterion that eludes quantitative benchmarking is the complexity of programming circuitry for synaptic memory. Significant infrastructure for producing programming signals could limit scalability. For example, floating-gate synapses often require programming signals at significantly higher voltages than are likely to be used in other parts of the network. For large-scale systems, memories with simple programming requirements will be at an advantage. Superconducting loop memory (Sec.\,\ref{Loops}) is intriguing from this standpoint, as the plasticity circuits operate with nearly identical signals and circuit blocks as those found in the rest of the network.

\subsection{Proposed Technologies}\label{Proposed}

\subsubsection{Room-temperature Analog Memories}
Many technologies have been proposed to implement synaptic weighting for room-temperature neuromorphic hardware, each with strengths and weaknesses \cite{upadhyay2019emerging}. The quest to find a suitable device for local synaptic memory dates back to the origins of the field, when Mead and colleagues investigated floating gate transistors \cite{diorio1998floating}. Since then, floating gate synapses have been used to implement STDP \cite{ramakrishnan2011floating}, are attractive as a mature alternative to emerging devices, and have been proposed for use in large-scale systems \cite{hasler2013finding}. However, there are concerns about high programming voltages, speed, and endurance that may limit floating-gate memories to situations with less-frequent updates. More recently, momentum has shifted to other technologies \cite{zahoor2020resistive}.  Memristive devices \cite{stsn2008,yast2012,ab2018}, commonly used in resistive random-access memory have emerged as a popular alternative, with recent demonstrations including monolithic integration with CMOS \cite{yin2019monolithically} and unsupervised pattern recognition with a simple network of synapses \cite{ielmini2018brain}. Questions remain about high variability (both cycle-to-cycle and device-to-device) \cite{dalgaty2019hybrid}, linearity, and endurance \cite{zahoor2020resistive}. Phase-change memory is another option, with its own demonstration of STDP \cite{ambrogio2016unsupervised}. Thermal management and endurance have been raised as issues \cite{upadhyay2019emerging, zahoor2020resistive}. Ferroelectric transistors present another alternative, as they have low variability, good potential for CMOS integration, and linearity \cite{kim2019ferroelectric}.  Spin-torque memory, 2D materials, and organic electronics have also been proposed as solutions. Interested readers should consult one of the many review articles on this topic \cite{kim2018recent, upadhyay2019emerging, zhang2020brain}. The field is burgeoning with new devices for synaptic memory, but to-date none has been dominant enough to monopolize research. To our knowledge, no technology has been able to simultaneously meet the targets in table\,\ref{tab:memory_metrics}, but progress in this area is encouraging.

\subsubsection{Superconducting Technologies}
Many of the aforementioned technologies may also apply to superconducting optoelectronic systems, but their cryogenic operation has been scarcely explored. Two other types of memory, only accessible at low temperatures, have received the most attention for superconducting systems: magnetic Josephson junctions (MJJs) and superconducting loop memories. An important distinction from room-temperature technologies is that for superconducting memory to be truly non-volatile, it must retain its state both in the absence of a power supply and upon warming to room-temperature.

\subsubsection{Magnetic Josepson Junctions}
MJJs have been proposed as a (nearly) non-volatile memory technology for superconducting neuromorphic computing. A two-terminal device, the critical current of an MJJ can be programmed by changing the magnetic order of a ferromagnetic material placed in the tunneling barrier of a JJ \cite{schneider2018ultralow}. MJJs are non-volatile with respect to electrical power, and there is optimism they can be made to retain their memory through a warm-up to room-temperature. Additionally, they provide remarkable performance with respect to the metrics given in Table \ref{tab:memory_metrics}. The energy per update is on the order of femtojoules (including cooling overhead), switching speeds are commensurate with firing rates exceeding 100\,GHz, and devices can be scaled to tens of nanometers. All of these metrics surpass the requirements for optoelectronic networks, and can be exploited in all-electronic superconducting networks as well \cite{schneider2018tutorial}. More work is needed to analyze the scaling potential of MJJs with respect to yield. The magnetic fields used during programming can be produced with magnetic control lines, but spin-torque mechanisms may provide a more scalable solution. Finding an efficient, scalable solution to programming MJJs in large-scale systems thus remains an area of research that will be critical to their potential for adoption.

\subsubsection{Loop Memory}\label{Loops}
Superconducting loop memories have been in use for decades by the superconducting electronics community \cite{vatu1998,ka1999}, but are not ideal for dense memory arrays commonly utilized as RAM in digital computing due to area concerns. In the case of optoelectronic spiking neural systems considered here, the objective is not to produce large RAM arrays, and size as well as addressing challenges do not emerge as significant impediments. Therefore, straightforward extensions of binary loop memories are the synaptic memory technology that appears most promising for the SOENs platform \cite{sh2018,shainline2019superconducting}. In these memory cells, circulating current persists indefinitely in a loop of superconducting wire. The current in the loop can be controlled by adding/removing magnetic-flux quanta with standard JJ circuitry. This memory loop is then inductively coupled to a wire supplying a bias current to a Josephson junction at the synapse ($J_{\mathrm{sf}}$ in Fig.\, \ref{fig:sup_synapse}(a)). When the synaptic SNSPD detects a photon, the biased junction will add an integer number of fluxons to another integrating superconductive loop (analogous to the membrane capacitance of a neuron). The number of fluxons added to the integration loop is a function of the bias supplied to the JJ, which is determined by the magnitude of current circulating in the memory loop. The number of analog memory levels in the memory loop is determined by the inductance of the loop, which is easily set with the length of a wire. High-kinetic-inductance materials \cite{tobo2018} enable memory storage loops with over a thousand levels (10 bits) to be fabricated in an area of 5\,\textmu m $\times$ 5\,\textmu m. 

The loop-memory approach has several strengths. The memory is nearly analog and updates are nearly linear. Memory is updated by the switching of a JJ, which involves only a change of the phase of the superconducting wave function. This phase can switch $10^{11}$ times in a second, so the endurance metric defined in the previous section is not an issue. This stands in contrast to room-temperature memories requiring material changes (filament formation, phase changes, etc.) which are often associated with degradation over time. Loop memory is also attractive from a fabrication perspective as it requires no additional materials or devices. The simplicity of the memory lends itself favorably to 3D integration, provided cross-talk from nearby loops can be mitigated. Plasticity circuits based on loop memories will also operate at the energy scale of single photons and flux quanta ($10^{-19}$\,J), which is commensurate with the rest of the circuitry in the network. This allows weight updates to be performed with the spikes the network produces in standard operation, reducing peripheral circuitry. There is no need to engineer differently shaped pulses for READ and WRITE operations, and the synapse does not need to be taken offline during programming. Simulations have demonstrated STDP learning with circuits containing four additional Josephson junctions \cite{shainline2019superconducting}. 

Two aspects of loop memory are concerning. First, loop memory is not strictly non-volatile. While circulating current can persist in a superconducting loop without any power supply, superconductivity must be maintained. If the temperature of the system is raised above the critical temperature of the superconducting material, the memory will be lost. Mechanisms for transferring weights stored in current loops to non-volatile solutions will need to be developed if the system's state is to be persevered upon reaching room-temperature (i.e. for maintenance or during a power interruption). The second weakness of loop memory is the size. The employed superconducting loops, as well as the transformers that couple them, will be large compared to all of the other solutions discussed. The consequences of these large-area components must be considered in the context of the entire system, which we discuss next.

\section{\label{sec:instantiation}System Level Considerations}
Here we consider aspects concerning the integration of the components previously discussed and how systems may reach the scale of the brain. Basic graph theory metrics and the assumption of 300-mm fabrication processes allow us to assess area constraints and the benefits of 3D integration. It is found that at least five planes of photonic routing will be required in either platform to achieve brain-scale systems. Prospects for 3D integration of active elements are addressed. It also must be stressed that an optoelectronic system of the complexity of the human brain will be abjectly impossible on a single 300-mm wafer in either case. A possible vision for connecting many wafers is discussed. Finally, we analyze cooling and power concerns, finding that neither should preclude the development of brain-scale systems in either platform.

\subsection{Considerations from Graph Theory}
Neurons in brain regions active in cognition, such as the cerebral cortex and hippocampus, are characterized by a high degree of connectivity\textemdash often in excess of ten thousand connections per neuron \cite{brsc1998,bu2006}. These connections often extend across appreciable spatial distances. Creating and maintaining these connections comes with high metabolic and spatial costs. The severely constrained biological brain would not support such expenditures if they were not advantageous to cognition \cite{busp2012}.

One reason why such high connectivity is necessary relates to efficient communication across the network. Rapid communication can only be achieved if the average path length across the network is small. In the language of graph theory, a network is a collection of nodes connected by edges. To calculate the shortest average path length across the network, one calculates the number of edges that must be traversed to travel from one node to another node in the network. One takes the mean of this quantity over all pairs of nodes. The shortest average path length ($\bar{L}$) is a global metric that offers a glimpse at the efficiency with which information can be communicated across space.

Equation \ref{eq:degree} provides the relationship between $\bar{L}$ and the number of edges connected to a node, or in our case, the number of synapses per neuron ($\bar{k}$) for a random network. In a random network, nearby and distant connections are equally probable. Specifically, the equation holds for Erd\"{o}s-R\'{e}nyi random graphs of networks with $N_{\mathrm{tot}}$ neurons \cite{frfr2004}:
\begin{equation}
\label{eq:degree}
\bar{k} = \mathrm{exp} \left[ \frac{\mathrm{ln}(N_{\mathrm{tot}})-\gamma}{\bar{L}-1/2} \right],
\end{equation}
where $\gamma \approx 0.5772$ is Euler's constant. For a network with $10^6$ neurons, each neuron must make nearly 10,000 connections to support an average path length of two, and 200 synapses must be formed to support a path length of three. For a network with $10^8$ neurons, more than 100,000 synapses are required for a path length of two, and more than 1,000 for a path length of three. The human hippocampus is a module with roughly $10^8$ neurons, each with 10,000-50,000 nearly spatially random connections. The objective of achieving an average path length between two and three may be an important reason why the hippocampus prioritizes this exceptional degree of connectivity \cite{bu2006}. The cerebral cortex in the human brain contains more than $10^{10}$ neurons, each with roughly 10,000 connections. This analysis indicates that a path length between two and three cannot be achieved across the entire cortex, and accordingly the cortex is constructed with a hierarchical, modular architecture \cite{si1962, mela2010} with high connectivity and efficient communication within smaller modules, and more sparse connectivity between modules separated by larger distances \cite{mo1997,mela2010,bosp2015,beba2017}.

While more sophisticated graph metrics can further elucidate the network concepts underlying cognition \cite{busp2009}, the simple, global metric of average shortest path length can help inform scaling analysis of artificial cognitive hardware at this early stage of development. We next consider the constraints $\bar{L}$ puts on the size of synaptic circuits.

\subsection{Generic Spatial Constraints}
\label{sec:spatial_constraints}
Based on the significance of the interplay between the hippocampus and cerebral cortex in cognition \cite{frbu2016}, we assume hardware for artificial neural systems will make use of similar architectural principles. Here we assume optoelectronic circuits will be fabricated using the conventional sequential, planar processing techniques of the silicon microelectronics industry. Photonic planes will implement the passive optical interconnects and electronic planes will accommodate all active electronics for neuronal function. We further specify to consideration of 300-mm wafers and seek a relationship between the network path length and the size of components on the wafer.

The area of a neuron occupied by its photonic waveguides can be approximated in a similar manner to the wires for electronic circuits \cite{ke1982}. This gives the following expression for the area of passive photonic circuitry:
\begin{equation}
\label{eq:area_p}
A_p = \left( \frac{k w_{\mathrm{wg}}}{p_p} \right)^2.
\end{equation}
$p_p$ is the number of photonic waveguide planes, $k$ is the degree of each neuron (assumed identical), and $w_{\mathrm{wg}}$ is the pitch of waveguides. The area of a neuron due to electronic synaptic circuits is given by
\begin{equation}
\label{eq:area_e}
A_e = \frac{k w_{\mathrm{sy}}^2}{p_e}.
\end{equation}
$w_{\mathrm{sy}}$ is the width of a synapse and $p_e$ is the number of planes of electronic circuits. Both $N_\mathrm{tot}A_p$ and $N_\mathrm{tot}A_e$ are subject to the area constraint of a 300-\,mm wafer. We use these relations to calculate the number of planes (electronic and photonic) that will be required to maintain a path length of 2.5 across a network of a given size (Fig.\,\ref{fig:num_planes}). See Appendix \ref{apx:scaling} for analysis of path length dependence on $w_{\mathrm{sy}}$ and $w_{\mathrm{wg}}$. A specific routing scheme is analyzed in reference \cite{shainline2019superconducting}. More than ten million neurons (less than a mouse brain) on a single 300-mm wafer appears out of reach for any platform.

\begin{figure}
    \centering
    \includegraphics[width=8.6cm]{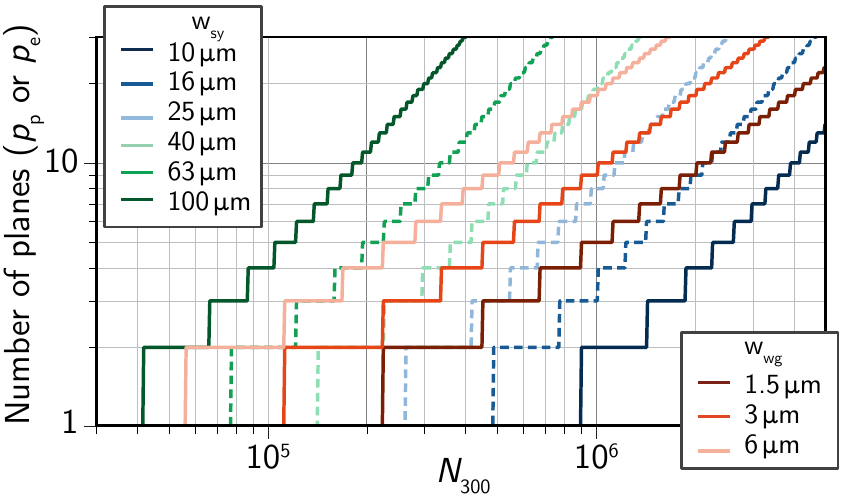} 
    \caption{Number of planes of active synaptic circuits ($p_e$) and passive photonic waveguides ($p_e$) required to maintain a path length of 2.5 as a function of the number of neurons on a 300-mm wafer ($N_{300}$).}
    \label{fig:num_planes}
\end{figure}

\subsection{Fabrication Processes}
\label{sec:fabrication}
We assume 300-mm silicon wafer processing. Wafer-scale integration has already been demonstrated for electronic neuromorphic systems \cite{schemmel2010wafer}. Still, even at this scale, reaching $10^6$ optoelectronic neurons per wafer is a tall order for either platform (Fig.\,\ref{fig:num_planes}). We choose this integration metric somewhat arbitrarily; $10^6$ neurons per wafer corresponds to $10^4$ wafers for a human-cortex-scale system. This is roughly the same order as the number of processing units in modern supercomputers. If this target is to be reached, 3D integration at some level will be necessary. From Fig.\,\ref{fig:num_planes}, it is clear that either platform will require a minimum of five photonic planes. Fortunately, photonic planes are quite amenable to 3D integration. Common waveguide materials include amorphous silicon (aSi), silicon nitride (SiN$_x$) and silicon oxynitride (SiO$_x$N$_y$). These dielectric materials can be deposited at low temperature, enabling several multi-planar demonstrations \cite{shpa2015,sahu2015,chbu2017,zhli2018}. Additionally, low-temperature deposition makes such processes compatible with back-end CMOS fabrication. It should be noted that five photonic planes represents a best-case scenario, as wider waveguides have lower loss and only minimal reduction in average path length (Appendix \ref{apx:scaling}).

3D integration of active electronics is less straightforward, particularly for the semiconductor approach. 3D CMOS integration has been the subject of decades of research \cite{ro1983,knan2008,saan2008,viba2011,zhxi2015,li2013,elfe2016,lish2017} and still faces uncertainty. Required high-temperature processing steps for dopant activation and contact anneals typically have a degrading effect on previous layers. Much of 3D integration of silicon microelectronics takes place at the die scale \cite{elfe2016}, which is incommensurate with the scale of systems under consideration. For the semiconductor scenario, the best course of action may be to abandon 3D active electronics altogether in favor of simply reducing the footprint ($w_{\mathrm{sy}}$) of synapses. We see again from Fig.\,\ref{fig:num_planes} that nearly $10^6$ neurons can be integrated on a single plane if each synapse is on the order of 10\,\textmu m\,$\times$\,10\,\textmu m. This may be a challenging benchmark to reach with high-functionality synapses implementing complex plasticity and dynamics. Subthreshold circuits that have embraced larger CMOS nodes for decreased variability may need to adjust to more modern nodes, of which there is some precedent \cite{rupa2019}. Additionally, photodetectors will be on the micron scale and long time-constant capacitors can require significant area (Appendix \ref{apx:TimeCon}) \cite{indiveri2019importance}. Both of these elements would however be fabricated on separate planes from MOSFETs.

Superconducting platforms would likely take the opposite approach, embracing 3D integration in the face of necessarily large device areas. Superconducting electronics, including active JJs, are routinely deposited at low temperatures ($<$\,180\,\textcelsius). Integrated circuits with two stacked planes of JJs have been demonstrated by two research laboratories \cite{tobo2019,anna2017}, along with multiple of planes of SNSPDs \cite{vema2012}. This is particularly important, as superconducting systems will not be able to reach $10^6$ neurons per wafer without 3D integration. A reasonable estimate for a superconducting synapse may be 30\,\textmu m on a side (Appendix \ref{apx:squid_area}). Such a size would require eight electronic planes. 

We note that even if $p_p = p_e = 1$, it is still possible to fabricate wafers with $10^6$ neurons, provided $\bar{k} = 100$, giving $\bar{L} = 3.5$ (Figs.\,\ref{fig:degree} and \ref{fig:path_length__vs__width} in Appendix \ref{apx:scaling}). While this does not match the short path lengths of cognitive circuits in the brain, such a network is likely to have significant technological and scientific utility while offering an intermediate-term practical objective.

\subsection{Constructing Multi-Wafer Systems}
Given that neither system will scale to billions of neurons on a single wafer, many wafers ($\sim$10,000) will need to be connected together to support human-brain-scale computing. A vision for a multi-wafer system is discussed in reference \cite{shainline2020optoelectronic} for the SOENs platform. Briefly, wafers are stacked and free-space optical communication is used to form highly inter-connected columns mimicking the modular structure of biological circuits \cite{mo1978,mo1997,mela2010,bosp2015,beba2017}. Columns are coupled to each other with lateral inter-wafer connections, but such connectivity is more sparse than that within a column. Optical fibers provide low-loss communication over long distances.

Achieving systems of this scale requires advances, particularly in wafer-scale circuit integration and system-level construction. A phenomenon akin to Moore's law, with ever-decreasing feature sizes enabling ever-higher integration density is unlikely to carry this concept forward, as many device sizes are limited by other physical considerations. Metrics related to number of planes of integrated circuits and number of wafers in a system may be more relevant to chart progress in neuromorphic supercomputing. Gradual progress may be possible by consistently scaling up, but it is difficult to envision this sustained trend without a powerful economic drive.  

\subsection{Power Consumption and Cooling}
\subsubsection{Cooling Systems}
Cooling systems will be a key component to either platform. For superconducting electronics, the system will fail completely if the temperature rises above the critical temperature ($T_c$). Superconducting neuromorphic systems will rely on niobium ($T_c = 9.3$K) or a material with a similarly low $T_c$. Liquid helium (4.2\,K) is the cryogen of choice for such temperatures. Cooling systems will add significantly to the power consumption of superconducting electronics. The power efficiency of a refrigeration system is measured by its specific power \cite{alekseev2015basics}. The specific power gives the number of watts consumed by the refrigeration system for every watt of heat removed. The theoretical limit for specific power, given by the Carnot limit, is $\frac{T_H - T_C}{T_C}$. For liquid helium temperature (4.2\,K), the Carnot limit demands that at least $74$\,watts of refrigeration power are required to remove every watt of heat produced on-chip if the system is operated in a 300\,K ambient. State-of-the-art systems have reached specific powers below $400$\,W/W. Auspiciously, the most efficient refrigeration systems also tend to have the highest heat loads. The ability to cool heat loads as high as 10\,kW at 4.4K have already been demonstrated by commercially available systems \cite{holmes2013energy}. Throughout this paper we assume a more conservative specific power of $1000$\,W/W, representative of the smaller scale cryogenic systems used in most laboratories today. It does not appear that cryogenic capability will be an insurmountable obstacle towards large-scale superconducting neural systems.

\subsubsection{Power Limitations}
\begin{figure}[h!]
    \centering
    \includegraphics[scale=1]{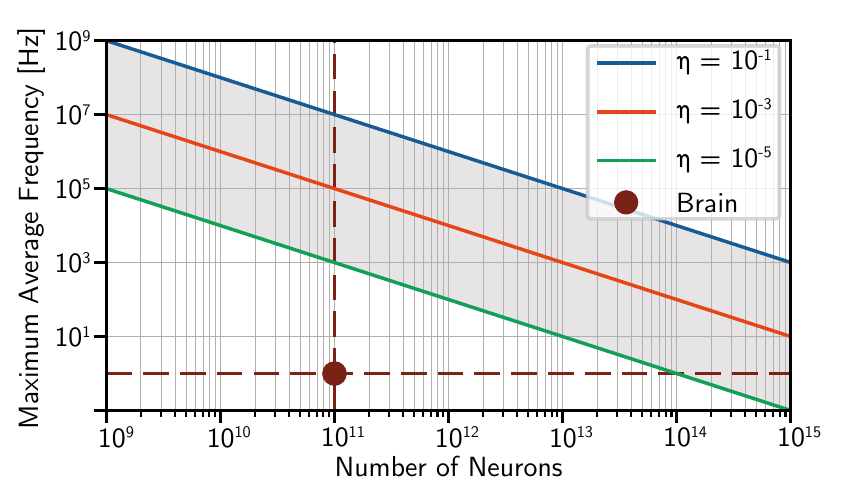}
    \caption{Tradeoff between size and average spiking frequency for a population of optoelectronic neurons with a power budget of 10 MW ($f = \frac{\mathrm{10\,MW}}{N_\mathrm{neurons}N E_{opt}}$). Fan-out ($N$) is $10^3$ and the optical energy needed at each synapse is assumed to be 1\,fJ (accounting for cooling in superconductor case). This likely would correspond to the limits of either superconductor or semiconductor neurons.}
    \label{fig:freq_size}
\end{figure}

Modern supercomputers typically consume megawatts of power. Tianhe 2, for instance, requires 17.8\,MW for operation (and another 6.4\,MW for cooling) \cite{tolpygo2016superconductor}. If we thus assume a total power budget of 10\,MW, we can analyze the trade-off between average firing rate and number of neurons. We assume 1\,fJ of optical energy is required to initiate a firing event at each synapse and plot the maximum average frequency spiking frequency for several different optical link efficiencies in Fig.\,\ref{fig:freq_size}.

Power does not appear to be a limiting factor in achieving brain-scale and brain-speed optoelectronic networks. If the power resources of modern supercomputers were dedicated to a brain-scale optoelectronic neuromorphic system, average spiking rates on the order of 10\,kHz ($10^{4}$ speedup over biology) appear feasible even with relatively inefficient optical links. Such a system may enable brain-scale computation with time accelerated by four orders of magnitude.

Another factor to consider is power density. There is a maximum power density that can be handled by heat removal systems for both the semiconducting and superconducting case. In the semiconductor case, high-performance computing routinely generates power densities of hundreds of watts per square centimeter \cite{tolpygo2016superconductor}. A theoretical limit of around 1\,kW/cm$^2$ is postulated in Ref.\,\onlinecite{zhirnov2003limits}. In contrast, superconducting systems will be required to operate at significantly lower power densities. Roughly 1\,W/cm$^2$ is a conservative limit for on-chip power density that can be cooled with liquid helium \cite{tolpygo2016superconductor}. Superconducting optical links appear to be capable of dissipating about three orders of magnitude less energy per bit, approximately cancelling out the limited power density requirements of superconducting systems in comparison with semiconductors. In practice, it might well be the case that mature, sophisticated synapses and neurons will occupy so much area that these power density limitations will be of no consequence. For instance, even with link efficiency of $\eta = 10^{-4}$, a synapse would require a lateral dimension of less than 30\,\textmu m for power density considerations to limit spiking to less than 1\,GHz. Section \ref{sec:instantiation} argued that superconducting synapses are not likely to be smaller than this. 10\,\textmu m semiconducting synapses could reach 1\,GHz with $1 \times 10^{-3}$ efficiency. However, optoelectronic systems will have nonuniform power dissipation across the chip/wafer, with most of the power being dissipated at the light sources. A more in-depth analysis is required to see if heat removal will be an issue near the light sources in particular, but for the superconducting case it is convenient that the light sources themselves are not superconducting, and can afford to be raised to higher temperatures without failure. Concerns about local heating may be assuaged with layouts that sufficiently shield and/or separate thermally sensitive devices from the light sources.

\section{\label{sec:conclusion}Conclusion}
The prospects of neuromorphic systems at the scale of the brain and beyond are tantalizing. The fan-out capability of optical communication coupled with the computational power of electronic circuitry makes optoelectronic systems a promising framework for realizing these high ambitions. However, there is no technology platform that is ready to support optoelectronic spiking networks of the scale and sophistication of the human brain. Making this vision a reality will require breakthroughs at the device level, no matter which path is chosen, particularly with regard to integrated light sources. Beyond that, several different classes of devices must be integrated alongside each other, which further reduces the likelihood for success. Efficient, densely integrated light sources, waveguide-coupled detectors, local memory devices, and capable neuronal circuitry all must be consolidated onto a single platform. Candidates for all requisite devices can be proposed for either semiconducting or superconducting platforms, and the two systems may be capable of similar performance. However, the technological paths toward achieving brain-scale systems with the two platforms diverge in important respects.

\begin{figure}[!h]
    \centering
    \includegraphics[width = 8.6cm]{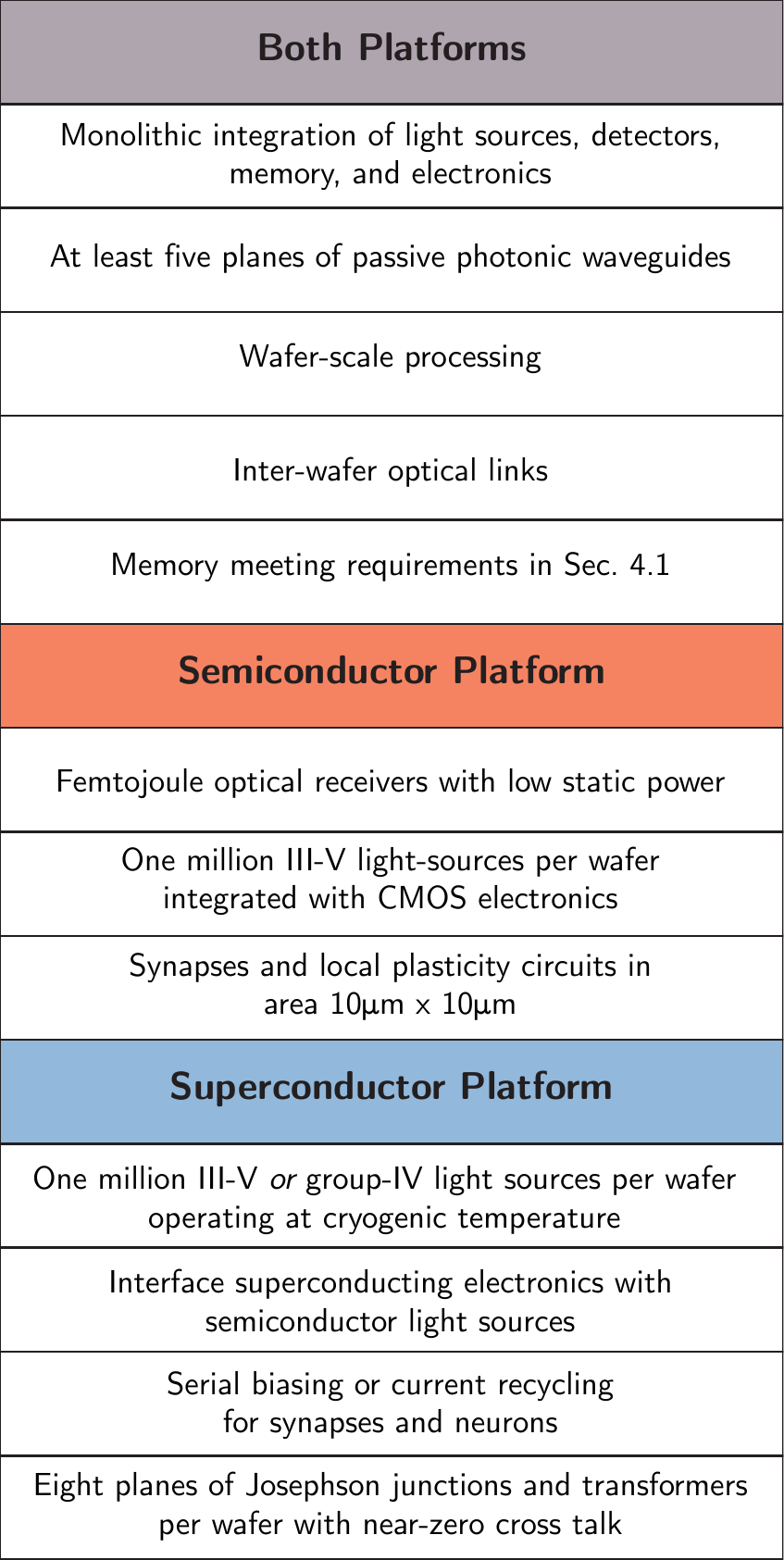}
    \caption{Summary of necessary hardware demonstrations for each platform if human-brain-scale artificial cognition is to be achieved.}
    \label{fig:targets}
\end{figure}

Semiconductor platforms hold advantages in technological maturity, room-temperature operation, and perhaps speed. Spike rates in excess of 10\,GHz may be feasible, but only for systems significantly smaller than the human brain due to power constraints. Semiconductor receivers can potentially operate with extremely low energies per spiking event (sub femto-joule), making them a worthy competitor of superconducting single photon detectors. However, these low energy receivers require significant optical power from integrated light sources. To achieve biological-scale fan-out, either very bright light sources, repeatering schemes (costing area and yield), or additional gain stages (costing power) will need to be included. In terms of neuronal computation, semiconductor neurons have already demonstrated impressive functionality and low-power operation that should be capable of integration with optical communication infrastructure, provided the long-standing challenges with CMOS-integrated III-V light sources can be overcome. Synaptic memory is a major open question, but a variety of non-volatile memory solutions have seen extensive investigation, and time will tell if one technology can meet the requirements we have laid out for brain-scale optoelectronic systems. 3D integration of transistors, photodetectors, and memory may not be a feasible solution, meaning aggressive scaling of synaptic circuits while maintaining complex functionality is perhaps a better strategy. The fabrication processes for mature semiconductor neural systems may prove to be prohibitively complicated and heterogeneous, perhaps requiring different processing strategies for sources, detectors, and memories. If wafer-scale monolithic integration of these components cannot be achieved, and chip-scale die-stacking techniques are required, the outlook for achieving brain-scale systems is limited.

Superconducting optoelectronic neural systems suffer from a comparatively primitive fabrication ecosystem, but the incorporation of superconducting devices provides several intriguing properties. SNSPD receivers place nearly the theoretical minimum burden on integrated light sources. This attribute compounds positively with the improvements in efficiency for light sources operating at cryogenic temperatures. Integration of light sources with superconducting electronics does not appear to have the same material integration challenges as integration with CMOS, but this state of affairs may be due to the lack of attention the effort has received. These factors make the large-scale integration of light sources appear more tractable than in the semiconductor case\textemdash perhaps even opening the door to silicon as an active optical material. Driving these light sources with superconducting electronics, however, has yet to demonstrate the performance required for this application. The implementation of a high-impedance pulse-and-reset circuit remains an open challenge. For computation, superconducting neuronal circuits appear just as capable of implementing complex neuronal and synaptic behaviors as their CMOS counterparts, but will need to be designed with serial biasing in order to scale. Additionally, some speed advantages present in superconducting electronics will be negated by the response time of SNSPDs ($<$1\,GHz ). Of course, even if maximum spike rates are limited to 20 MHz, this would still represent a speed-up of four orders of magnitude over biological systems. Memory seems to be a strength for the superconducting platform, as superconductivity provides new avenues of storing synaptic weights. Loop memory in particular may be capable of implementing plasticity mechanisms that operate with only the signals produced through normal network activity. Caution is in order here, however, as superconducting synaptic plasticity mechanisms have scarcely been explored. 3D integration may yield more readily in the superconductor platform. The inconvenience of cryogenic cooling is a serious consideration, but power and heat removal estimations indicate this is unlikely to be a limiting factor for brain-scale systems. \textit{If} all these issues can be resolved, superconducting optoelectronic systems may require simpler manufacturing processes than the semiconductor approach, as the material ecosystem could potentially be parsimonious. Of course, superconducting foundries are far less developed than their semiconductor counterparts, which may negate these advantages in the near-term.

We would be remiss to paint the quest for neuromorphic supercomputing as only a question of hardware. The inner workings of the brain are the subject of intense investigation, and the emergent phenomena of cognition and consciousness remain taunting, increasingly lonely enigmas entrenched in the netherworld between philosophy and science. Watershed breakthroughs in neuroscience and algorithmic development will be required for the discussed hardware platforms to have practical applications, although the hardware platforms themselves may be of use in helping to unravel some of these mysteries. The question of whether it is prudent to develop hardware before algorithms has pestered the field of neuromorphic computing since its inception. In this case, we believe that the length of development, rich opportunities for spin-off technologies, and inestimable potential make such hardware development well-worth pursuing even at this incipient stage.  

\section*{Acknowledgements}
We thank Dr. Brian Hoskins, Dr. Advait Madhavan, and Dr. Alexander Tait for helpful insights and conversation.
\appendix

\section{Implementing Long Time Constants}\label{apx:TimeCon}
\begin{figure}[!h]
    \centering
    \includegraphics[width=8.6cm]{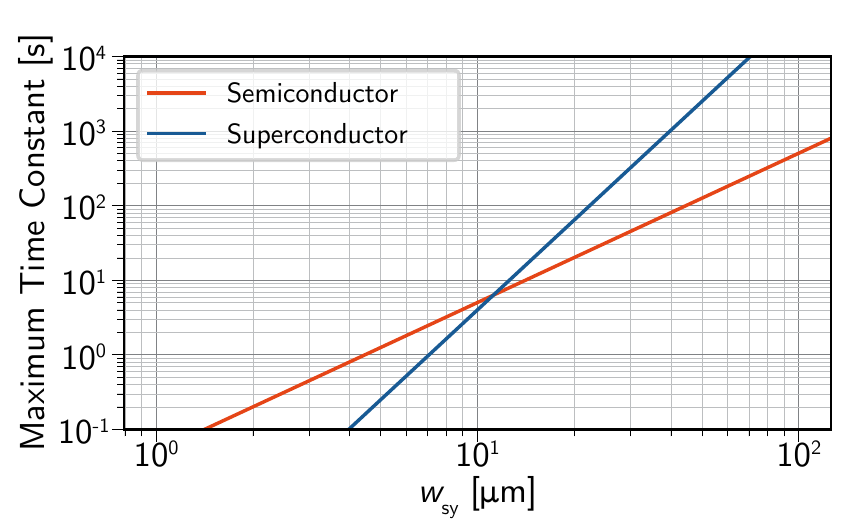}
    \caption{Maximum achievable time constant as a function of synapse width.}
    \label{fig:timecon}
\end{figure}

For the DPI synapse (Sec. 3) the time constant is given by \cite{chicca2014neuromorphic}:
\begin{equation}
    \tau = \frac{C_{\mathrm{si}}V_{\mathrm{th}}}{\kappa I_{\tau}},
\end{equation}
where $I_{\tau}$ is the current leaking off of the capacitor, set with transistor $T_3$, $V_{\mathrm{th}}$ is the thermal voltage, and $\kappa$ is the sub-threshold slope factor (typically order 1). Operating in the subthreshold regime allows $I_{\tau}$ to be reduced to femtoamps \cite{linares2003design}. Metal Insulator Metal (MIM) capacitors ultizing high-k dielectrics can reach capacitance densities around 20\,fF/\textmu m$^2$ \cite{wu2009metal}. The maximum achievable time constant as a function of synapse width is shown in figure \ref{fig:timecon} for $I_{\tau}$ = 10\,fA, $\kappa$ = 1, and $V_{\mathrm{th}}$ = 25\,mV. Since MIM capacitors can be fabricated on a separate layer from transistors, the entire 10\,\textmu m $\times$\,10\,\textmu m area per synapse target identified in Sec. \ref{sec:instantiation} could be dedicated to capacitor area.

For SOENs synapses, $L_{\mathrm{si}}/r_{\mathrm{si}}$ sets the time constant. Inductors and resistors will most likely be fabricated on separate layers, again conserving space. A meandering geometry gives the maximum inductance $L_{\mathrm{si}}$, that can be fabricated in an area $w_{sy}^2$ as:
\begin{equation}
    L_{\mathrm{si}} = \frac{w_{sy}^2L_{\Box}}{w_{\mathrm{wire}}(w_{\mathrm{wire}} + w_{\mathrm{gap}})},
\end{equation}
where $L_{\Box}$ is the inductance per square of the material, and $w_{\mathrm{wire}}$ and $w_{\mathrm{gap}}$ are the minimum feature sizes. Small value resistors are fabricated by putting many wide resistances in parallel. The smallest (nonzero of course) resistor that can be fabricated in an area of $w_{sy}^2$ is then:
\begin{equation}
    r_{\mathrm{si}} = \frac{R_sw_{\mathrm{gap}}(w_{\mathrm{wire}} + w_{\mathrm{gap}})}{w_{sy}^2},
\end{equation}
where $R_s$ is the sheet resistance. The maximum time constant ($\tau_{\mathrm{max}}$) in area $w_{sy}^2$ is given by $L_{\mathrm{si}}/r_{\mathrm{si}}$:
\begin{equation}
    \tau_{\mathrm{max}} = \frac{w_{sy}^4L_{\Box}}{R_sw_{\mathrm{wire}}w_{\mathrm{gap}}(w_{\mathrm{wire}}+w_{\mathrm{gap}})^2}
\end{equation}
The maximum achievable time constants as a function of synapse width is plotted in figure \ref{fig:timecon} for $R_s = .001$\,$\Omega/\square$ (corresponds to a 200\,nm gold layer with resistivity of $2\times 10^{-10}\,\Omega\cdot$m at 4\,K \cite{matula1979electrical}), $w_{\mathrm{wire}} = w_{\mathrm{gap}} = 100$\,nm, and $L_{\Box} = 160$\,pH$/\square$, corresponding to MoSi. For the large synaptic areas expected to be available to superconducting synapses via 3D integration, the superconducting approach can support significantly larger time constants than the semiconducting case. In Ref.\,\onlinecite{indiveri2019importance}, it is suggested that the human brain itself is limited by the maximum achievable time constant, and relies on network dynamics for certain types of long-term memory. The ramifications of hardware with time-constants far greater than biology are intriguing. 

\section{Area and Energy in Superconducting Synapses}\label{apx:squid_area}
JJs will never be as small as MOSFETs \cite{tolpygo2016superconductor}, but the area of synaptic circuits is not limited by JJs. Large inductors/transformers that couple flux from storage loops to receiving SQUIDs are likely to be the components that consume the most area. The size of these components is determined by the critical current of the junctions used in the SQUIDs ($I_c$), the permeability of free space ($\mu_0$), SQUID inductance ($L$), and the magnetic flux quantum ($\Phi_0$). Based on the typical SQUID design criterion \cite{clbr2006} we expect $2LI_c/\Phi_0 = 1$. For a simple washer-type inductor geometry, $L \approx 1.25\mu_0 w_{\mathrm{sq}}$, where $w_{\mathrm{sq}}$ is the inner dimension of the hole \cite{jaycox1981planar}. The energy, $E_{\mathrm{sq}}$ to produce two fluxons is approximately $2I_c\Phi_0$ for an appropriately biased junction. There is thus a trade-off between the size of a SQUID and its energy consumption:
\begin{equation}
    w_{\mathrm{sq}} \approx \frac{\Phi_0^2}{1.25\mu_0E_{\mathrm{sq}}}
\end{equation}

$I_c$ is further constrained by noise and the ability to interface with SNSPDs. $I_c = 300$\,\textmu A is reasonable and would correspond to $w_{\mathrm{sq}}\approx2.2$\,\textmu m and $E_{\mathrm{sq}} \approx 1.2$\,aJ. If the optical energy per synapse is around 100\,aJ for a 1\% efficient link, about 170 fluxons can be produced per synapse event without dominating the power budget. In practice, each SQUID may require a lateral dimension about five times larger than $w_\mathrm{sq}$ to account for washer width, wiring, and spacing to minimize cross-talk. Additionally, each synapse
will likely be composed of three or four SQUIDs\textemdash one for synaptic integration, one for loop memory, and perhaps two more for various plasticity functions. A reasonable estimate for synaptic size is then about 30\,\textmu m\,$\times$\,30\,\textmu m. 

\section{Further Scaling Analysis}\label{apx:scaling}
\begin{figure}[!h]
    \centering
    \includegraphics[width=8.6cm]{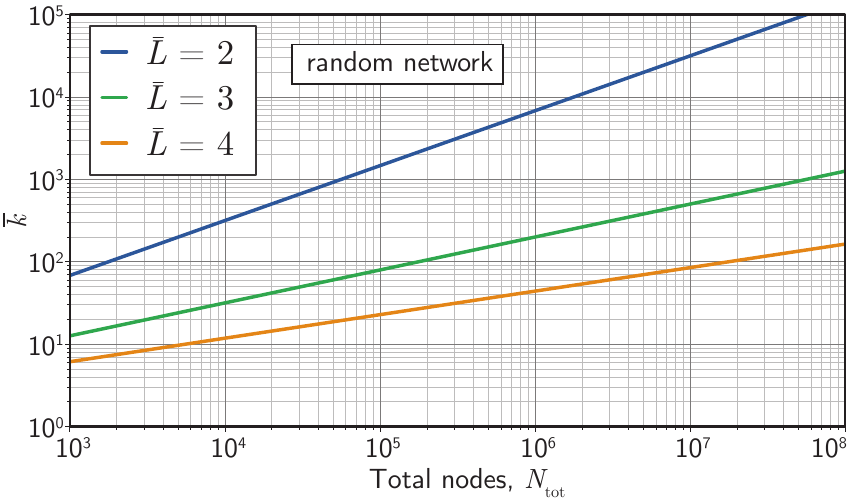}
    \caption{Average node degree as a function of network size for different path lengths.}
    \label{fig:degree}
\end{figure}

\begin{figure}[!h]
    \centering
    \includegraphics[width=8.6cm]{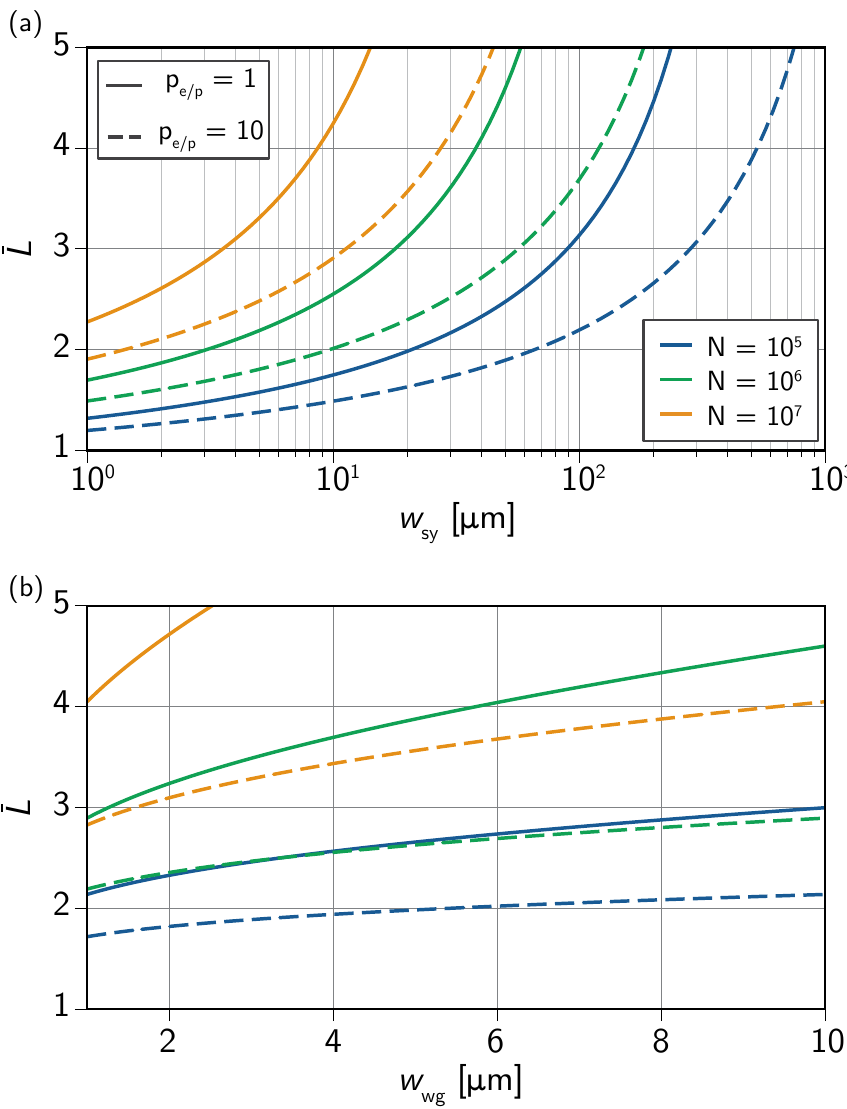}
    \caption{Path length versus feature size. (a) Path length versus width of electronic synapses. (b) Path length versus waveguide pitch. In both figures the solid lines correspond to a single plane of synapses or waveguides, and the dashed lines correspond to 10 planes of synapses or waveguides. The blue, green, and yellow traces correspond to the cases of $10^5$, $10^6$, and $10^7$ neurons per 300-mm wafer, respectively.}
    \label{fig:path_length__vs__width}
\end{figure}

Figures \ref{fig:degree} and \ref{fig:path_length__vs__width} provide insight into how network connectivity constrains hardware for any planar fully-dedicated system. Figure \ref{fig:degree} plots Eq.\,\ref{eq:degree}, giving the average node degree (number of synapses per neuron) necessary to maintain a given path length as a function of network size.

Figure \ref{fig:path_length__vs__width} plots the minimum achievable path length as function of synaptic width and waveguide width (Eqs.\,\ref{eq:area_e} and \ref{eq:area_p}). Path length is relatively insensitive to waveguide pitch, suggesting that wider, and therefore lower loss waveguides may be beneficial. In contrast, we see that the synaptic size ($w_{\mathrm{sy}}$) can be a major impediment to achieving low path lengths. 

\bibliographystyle{unsrt}
\bibliography{bib}

\begin{thebibliography}{100}

\bibitem{furber2016large}
Steve Furber.
\newblock Large-scale neuromorphic computing systems.
\newblock {\em Journal of neural engineering}, 13(5):051001, 2016.

\bibitem{shainline2019superconducting}
Jeffrey~M Shainline, Sonia~M Buckley, Adam~N McCaughan, Jeffrey~T Chiles, Amir
  Jafari~Salim, Manuel Castellanos-Beltran, Christine~A Donnelly, Michael~L
  Schneider, Richard~P Mirin, and Sae~Woo Nam.
\newblock Superconducting optoelectronic loop neurons.
\newblock {\em Journal of Applied Physics}, 126(4):044902, 2019.

\bibitem{young2019review}
Aaron~R Young, Mark~E Dean, James~S Plank, and Garrett~S Rose.
\newblock A review of spiking neuromorphic hardware communication systems.
\newblock {\em IEEE Access}, 7:135606--135620, 2019.

\bibitem{hennessy2011computer}
John~L Hennessy and David~A Patterson.
\newblock {\em Computer architecture: a quantitative approach}.
\newblock Elsevier, 2011.
\newblock Appendix F, pg. F-53.

\bibitem{seda2016}
C.~Segal, A.~Dalakoti, M.~Miller, and F.~Brewer.
\newblock {Connectivity Effects on Energy and Area for Neuromorphic System with
  High Speed Asynchronous Pulse Mode Links}.
\newblock {\em SLIP}, 16, 2016.

\bibitem{marder1987neurotransmitters}
E~Marder.
\newblock Neurotransmitters and neuromodulators.
\newblock In {\em The crustacean stomatogastric system}, pages 263--306.
  Springer, 1987.

\bibitem{euler2001dendritic}
Thomas Euler and Winfried Denk.
\newblock Dendritic processing.
\newblock {\em Current opinion in neurobiology}, 11(4):415--422, 2001.

\bibitem{shainline2017superconducting}
Jeffrey~M Shainline, Sonia~M Buckley, Richard~P Mirin, and Sae~Woo Nam.
\newblock Superconducting optoelectronic circuits for neuromorphic computing.
\newblock {\em Physical Review Applied}, 7(3):034013, 2017.

\bibitem{shainline2019fluxonic}
Jeffrey~M Shainline.
\newblock Fluxonic processing of photonic synapse events.
\newblock {\em IEEE Journal of Selected Topics in Quantum Electronics},
  26(1):1--15, 2019.

\bibitem{sh2021}
J.M. Shainline.
\newblock {Optoelectronic Intelligence}.
\newblock {\em Appl. Phys. Lett.}, 118:160501, 2021.

\bibitem{miller2017attojoule}
David~AB Miller.
\newblock Attojoule optoelectronics for low-energy information processing and
  communications.
\newblock {\em Journal of Lightwave Technology}, 35(3):346--396, 2017.

\bibitem{spga2011}
J.P. Sprengers, A.~Gaggero, D.~Sahin, S.~Jahanmirinejad, G.~Frucci,
  F.~Mattioli, R.~Leoni, J.~Beetz, M.~Lermer, M.~Kamp, S.~H{\" o}fling,
  R.~Sanjines, and A.~Fiore.
\newblock Waveguide superconducting single-photon detectors for integrated
  quantum photonic circuits.
\newblock {\em Appl. Phys. Lett.}, 99:181110, 2011.

\bibitem{pesc2012}
W.~Pernice, C.~Schuck, O.~Minaeva, M.~Li, G.~Goltsman, A.~Sergienko, and
  H.~Tang.
\newblock High speed travelling wave single-photon detectors with near-unity
  quantum efficiency.
\newblock {\em Nat. Comm.}, 3:1325, 2012.

\bibitem{akhlaghi2015waveguide}
Mohsen~K Akhlaghi, Ellen Schelew, and Jeff~F Young.
\newblock Waveguide integrated superconducting single-photon detectors
  implemented as near-perfect absorbers of coherent radiation.
\newblock {\em Nature communications}, 6(1):1--8, 2015.

\bibitem{feka2015}
S.~Ferrari, O.~Kahl, V.~Kovalyuk, G.N. Goltsman, A.~Korneev, and W.H.P.
  Pernice.
\newblock Waveguide-integrated single- and multi-photon detecton at telecom
  wavelengths using superconducting nanowires.
\newblock {\em Appl. Phys. Lett.}, 106:151101, 2015.

\bibitem{saga2015}
D.~Sahin, A.~Gaggero, J.-W. Weber, I.~Agafonov, M.A. Verheijen, F.~Mattioli,
  J.~Beetz, M.~Kamp, S.~H{\" o}fling, M.C.M. van~de Sanden, R.~Leoni, , and
  A.~Fiore.
\newblock Waveguide nanowire superconducting single-photon detectors fabricated
  on gaas and the study of their optical properties.
\newblock {\em IEEE J. Sel. Top. Quant. Electron.}, 21:3800210, 2015.

\bibitem{shbu2017b}
J.M. Shainline, S.M. Buckley, N.~Nader, C.M. Gentry, K.C. Cossel, J.W. Cleary,
  M.~Popovi\'{c}, N.R. Newbury, S.W. Nam, and R.P. Mirin.
\newblock Room-temperature-deposited dielectrics and superconductors for
  integrated photonics.
\newblock {\em Opt. Express}, 25:10322, 2017.

\bibitem{ferrari2018waveguide}
Simone Ferrari, Carsten Schuck, and Wolfram Pernice.
\newblock Waveguide-integrated superconducting nanowire single-photon
  detectors.
\newblock {\em Nanophotonics}, 7(11):1725--1758, 2018.

\bibitem{buta2020}
S.M. Buckley, A.N. Tait, J.~Chiles, A.N. McCaughan, S.~Khan, R.P. Mirin, S.W.
  Nam, and J.M. Shainline.
\newblock {Integrated-Photonic Characterization of Single-Photon Detectors for
  Use in Neuromorphic Synapses}.
\newblock {\em Phys. Rev. Applied}, page 054008, 2020.

\bibitem{razavi2012design}
Behzad Razavi.
\newblock {\em Design of integrated circuits for optical communications}.
\newblock John Wiley \& Sons, 2012.

\bibitem{mave2013}
F.~Marsili, V.B. Verma, J.A. Stern, S.~Harrington, A.E. Lita, T.~Gerrits,
  I.~Vayshnker, B.~Baek, M.D. Shaw, R.P. Mirin, and S.W. Nam.
\newblock Detecting single infrared photons with 93\% system efficiency.
\newblock {\em Nat. Photon.}, 7:210, 2013.

\bibitem{rene2020}
D.V. Reddy, R.R. Nerem, S.W. Nam, R.P. Mirin, and V.B. Verma.
\newblock {Superconducting nanowire single-photon detectors with 98\% system
  detection efficiency at 1550\,nm}.
\newblock {\em Optica}, 7:1649, 2020.

\bibitem{stein2005neuronal}
Richard~B Stein, E~Roderich Gossen, and Kelvin~E Jones.
\newblock Neuronal variability: noise or part of the signal?
\newblock {\em Nature Reviews Neuroscience}, 6(5):389--397, 2005.

\bibitem{mcdonnell2011benefits}
Mark~D McDonnell and Lawrence~M Ward.
\newblock The benefits of noise in neural systems: bridging theory and
  experiment.
\newblock {\em Nature Reviews Neuroscience}, 12(7):415--425, 2011.

\bibitem{allen1994evaluation}
Christina Allen and Charles~F Stevens.
\newblock An evaluation of causes for unreliability of synaptic transmission.
\newblock {\em Proceedings of the National Academy of Sciences},
  91(22):10380--10383, 1994.

\bibitem{li1997}
J.E. Lisman.
\newblock Bursts as a unit of neural information: making unreliable synapses
  reliable.
\newblock {\em Trends Neurosci.}, 20:38, 1997.

\bibitem{vema2012}
V.B. Verma, F.~Marsili, S.~Harrington, A.E. Lita, R.P. Mirin, and S.W. Nam.
\newblock A three-dimensional, polarization-insensitive superconducting
  nanowire avalanche photodetector.
\newblock {\em Appl. Phys. Lett.}, 101:251114, 2012.

\bibitem{rosenberg2013high}
D~Rosenberg, AJ~Kerman, RJ~Molnar, and EA~Dauler.
\newblock High-speed and high-efficiency superconducting nanowire single photon
  detector array.
\newblock {\em Optics express}, 21(2):1440--1447, 2013.

\bibitem{vetter2016cavity}
Andreas Vetter, Simone Ferrari, Patrik Rath, Rasoul Alaee, Oliver Kahl, Vadim
  Kovalyuk, Silvia Diewald, Gregory~N Goltsman, Alexander Korneev, Carsten
  Rockstuhl, et~al.
\newblock Cavity-enhanced and ultrafast superconducting single-photon
  detectors.
\newblock {\em Nano letters}, 16(11):7085--7092, 2016.

\bibitem{wove2019}
E.E. Wollman, V.B. Verma, A.E. Lita, W.H. Farr, M.D. Shaw, R.P. Mirin, and S.W.
  Nam.
\newblock Kilopixel array of superconducting nanowire single-photon detectors.
\newblock {\em Opt. Express}, 27:35279, 2019.

\bibitem{debaes2003receiver}
Christof Debaes, Aparna Bhatnagar, Diwakar Agarwal, Ray Chen, Gordon~A Keeler,
  Noah~C Helman, Hugo Thienpont, and David~AB Miller.
\newblock Receiver-less optical clock injection for clock distribution
  networks.
\newblock {\em IEEE journal of selected topics in quantum electronics},
  9(2):400--409, 2003.

\bibitem{derose2011ultra}
Christopher~T DeRose, Douglas~C Trotter, William~A Zortman, Andrew~L Starbuck,
  Moz Fisher, Michael~R Watts, and Paul~S Davids.
\newblock Ultra compact 45 ghz cmos compatible germanium waveguide photodiode
  with low dark current.
\newblock {\em Optics express}, 19(25):24897--24904, 2011.

\bibitem{meor2014}
K.K. Mehta, J.S. Orcutt, J.M. Shainline, O.~Tehar-Zahav, Z.~Sternberg,
  R.~Meade, M.A. Popovi{\' c}, and Rajeev~J. Ram.
\newblock Polycrystalline silicon ring resonator photodiodes in a bulk
  complementary metal-oxide-semiconductor process.
\newblock {\em Opt. Lett.}, 39:1061, 2014.

\bibitem{zhang2020scalable}
Yu~Zhang, Anirban Samanta, Kuanping Shang, and SJ~Ben Yoo.
\newblock Scalable 3d silicon photonic electronic integrated circuits and their
  applications.
\newblock {\em IEEE Journal of Selected Topics in Quantum Electronics},
  26(2):1--10, 2020.

\bibitem{nozaki2018forward}
Kengo Nozaki, Shinji Matsuo, Takuro Fujii, Koji Takeda, Akihiko Shinya, Eiichi
  Kuramochi, and Masaya Notomi.
\newblock Forward-biased nanophotonic detector for ultralow-energy dissipation
  receiver.
\newblock {\em APL Photonics}, 3(4):046101, 2018.

\bibitem{assefa2010reinventing}
Solomon Assefa, Fengnian Xia, and Yurii~A Vlasov.
\newblock Reinventing germanium avalanche photodetector for nanophotonic
  on-chip optical interconnects.
\newblock {\em Nature}, 464(7285):80--84, 2010.

\bibitem{virot2014germanium}
L{\'e}opold Virot, Paul Crozat, Jean-Marc F{\'e}d{\'e}li, Jean-Michel Hartmann,
  Delphine Marris-Morini, Eric Cassan, Fr{\'e}d{\'e}ric Boeuf, and Laurent
  Vivien.
\newblock Germanium avalanche receiver for low power interconnects.
\newblock {\em Nature communications}, 5(1):1--6, 2014.

\bibitem{pizzone2020analysis}
Andrea Pizzone, Srinivasan~Ashwyn Srinivasan, Peter Verheyen, Guy Lepage,
  Sadhishkumar Balakrishnan, and Joris Van~Campenhout.
\newblock Analysis of dark current in ge-on-si photodiodes at cryogenic
  temperatures.
\newblock In {\em 2020 IEEE Photonics Conference (IPC)}, pages 1--2. IEEE.

\bibitem{mccaughan2019superconducting}
Adam~N McCaughan, Varun~B Verma, Sonia~M Buckley, JP~Allmaras, AG~Kozorezov,
  AN~Tait, SW~Nam, and JM~Shainline.
\newblock A superconducting thermal switch with ultrahigh impedance for
  interfacing superconductors to semiconductors.
\newblock {\em Nature electronics}, 2(10):451--456, 2019.

\bibitem{romeira2019physical}
Bruno Romeira and Andrea Fiore.
\newblock Physical limits of nanoleds and nanolasers for optical
  communications.
\newblock {\em Proceedings of the IEEE}, 108(5):735--748, 2019.

\bibitem{iyxi1993}
S.~S. Iyer and Y.-H. Xie.
\newblock Light emission from silicon.
\newblock {\em Science}, 260:40, 1993.

\bibitem{shxu2007}
J.M. Shainline and J.~Xu.
\newblock Silicon as an emissive optical medium.
\newblock {\em Laser and Photonics Reviews}, 1:334, 2007.

\bibitem{warga2008electroluminescence}
J~Warga, R~Li, SN~Basu, and L~Dal~Negro.
\newblock Electroluminescence from silicon-rich nitride/silicon superlattice
  structures.
\newblock {\em Applied Physics Letters}, 93(15):151116, 2008.

\bibitem{wabo2005}
Robert~J. Walters, George~I. Bourianoff, and Harry~A. Atwater.
\newblock Field-effect electroluminescence in silicon nanocrystals.
\newblock {\em Nature Materials}, 4:143--146, 2005.

\bibitem{enpo1985}
H.~Ennen, G.~Pomrenke, A.~Axmann, K.~Eisele, W.~Haydl, and J.~Schneider.
\newblock 1.54 $\mu$ m electroluminescence of erbium-doped silicon grown by
  molecular beam epitaxy.
\newblock {\em Appl. Phys. Lett.}, 46:381, 1985.

\bibitem{paga1996}
J.~Palm, F.~Gan, B.~Zheng, J.~Michel, and L.C. Kimerling.
\newblock Electroluminescence of erbium-doped silicon.
\newblock {\em Phys. Rev. B}, 54:17603, 1996.

\bibitem{brha1986}
T.G. Brown and D.G. Hall.
\newblock Observation of electroluminescence from excitons bound to
  isoelectronic impurities in crystalline silicon.
\newblock {\em J. Appl. Phys.}, 59:1399, 1986.

\bibitem{brbr1989}
P.L. Bradfield, T.G. Brown, and D.G. Hall.
\newblock Electroluminescence from sulfur impurities in an p-n junction formed
  in epitaxial silicon.
\newblock {\em Appl. Phys. Lett.}, 55:100, 1989.

\bibitem{rosh2007b}
E.~Rotem, J.M. Shainline, and J.M. Xu.
\newblock Electroluminescence of nanopatterned silicon with carbon implantation
  and solid phase epitaxial regrowth.
\newblock {\em Opt. Express}, 15:14099, 2007.

\bibitem{bata2007}
Jiming Bao, Malek Tabbael, Taegon Kim, Supakit Charnvanichborikarn, James~S.
  Williams, Michael~J. Aziz, and Federico Capasso.
\newblock Point defect engineered si sub-band gap light-emitting diode.
\newblock {\em Opt. Express}, 15:6727, 2007.

\bibitem{ng2001efficient}
Wai~Lek Ng, MA~Lourenco, RM~Gwilliam, S~Ledain, Guosheng Shao, and KP~Homewood.
\newblock An efficient room-temperature silicon-based light-emitting diode.
\newblock {\em Nature}, 410(6825):192--194, 2001.

\bibitem{kvba2004}
V.~Kveder, M.~Badylevich, E.~Steinman, A.~Izotov, M.~Seibt, and W.~Schr{\"
  o}ter.
\newblock Room-temperature silicon light-emitting diodes based on dislocation
  luminescence.
\newblock {\em Appl. Phys. Lett.}, 84:2106, 2004.

\bibitem{grzh2001}
Martin~A. Green, Jianhua Zaho, Aihua Wang, Peter~J. Reese, and Michael Gal.
\newblock Efficient silicon light-emitting diodes.
\newblock {\em Nature}, 412:805, 2001.

\bibitem{sun2009toward}
Xiaochen Sun, Jifeng Liu, Lionel~C Kimerling, and Jurgen Michel.
\newblock Toward a germanium laser for integrated silicon photonics.
\newblock {\em IEEE Journal of Selected Topics in Quantum Electronics},
  16(1):124--131, 2009.

\bibitem{ishikawa2003strain}
Yasuhiko Ishikawa, Kazumi Wada, Douglas~D Cannon, Jifeng Liu, Hsin-Chiao Luan,
  and Lionel~C Kimerling.
\newblock Strain-induced band gap shrinkage in ge grown on si substrate.
\newblock {\em Applied Physics Letters}, 82(13):2044--2046, 2003.

\bibitem{ghrib2012control}
A~Ghrib, M~De~Kersauson, M~El~Kurdi, R~Jakomin, G~Beaudoin, S~Sauvage,
  G~Fishman, G~Ndong, M~Chaigneau, R~Ossikovski, et~al.
\newblock Control of tensile strain in germanium waveguides through silicon
  nitride layers.
\newblock {\em Applied Physics Letters}, 100(20):201104, 2012.

\bibitem{tani2021enhanced}
Kazuki Tani, Katsuya Oda, Momoko Deura, and Tatemi Ido.
\newblock Enhanced room-temperature electroluminescence from a germanium
  waveguide on a silicon-on-insulator diode with a silicon nitride stressor.
\newblock {\em Optics Express}, 29(3):3584--3595, 2021.

\bibitem{liu2007tensile}
Jifeng Liu, Xiaochen Sun, Dong Pan, Xiaoxin Wang, Lionel~C Kimerling, Thomas~L
  Koch, and Jurgen Michel.
\newblock Tensile-strained, n-type ge as a gain medium for monolithic laser
  integration on si.
\newblock {\em Optics express}, 15(18):11272--11277, 2007.

\bibitem{el2009enhanced}
M~El~Kurdi, T~Kociniewski, T-P Ngo, J~Boulmer, D~Debarre, P~Boucaud,
  JF~Damlencourt, O~Kermarrec, and D~Bensahel.
\newblock Enhanced photoluminescence of heavily n-doped germanium.
\newblock {\em Applied Physics Letters}, 94(19):191107, 2009.

\bibitem{sun2009direct}
Xiaochen Sun, Jifeng Liu, Lionel~C Kimerling, and Jurgen Michel.
\newblock Direct gap photoluminescence of n-type tensile-strained ge-on-si.
\newblock {\em Applied Physics Letters}, 95(1):011911, 2009.

\bibitem{camacho2013direct}
Rodolfo Camacho-Aguilera, Zhaohong Han, Yan Cai, Lionel~C Kimerling, and Jurgen
  Michel.
\newblock Direct band gap narrowing in highly doped ge.
\newblock {\em Applied Physics Letters}, 102(15):152106, 2013.

\bibitem{virgilio2013radiative}
Michele Virgilio, CL~Manganelli, Giuseppe Grosso, G~Pizzi, and G~Capellini.
\newblock Radiative recombination and optical gain spectra in biaxially
  strained n-type germanium.
\newblock {\em Physical review B}, 87(23):235313, 2013.

\bibitem{sun2009room}
Xiaochen Sun, Jifeng Liu, Lionel~C Kimerling, and Jurgen Michel.
\newblock Room-temperature direct bandgap electroluminesence from ge-on-si
  light-emitting diodes.
\newblock {\em Optics letters}, 34(8):1198--1200, 2009.

\bibitem{liu2010ge}
Jifeng Liu, Xiaochen Sun, Rodolfo Camacho-Aguilera, Lionel~C. Kimerling, and
  Jurgen Michel.
\newblock Ge-on-si laser operating at room temperature.
\newblock {\em Opt. Lett.}, 35(5):679--681, Mar 2010.

\bibitem{Fadaly2020}
Elham~M.T. Fadaly, Alain Dijkstra, Jens~Ren{\`{e}} Suckert, Dorian Ziss,
  Marvin~A.J. van Tilburg, Chenyang Mao, Yizhen Ren, Victor~T. van Lange,
  Ksenia Korzun, Sebastian K{\"{o}}lling, Marcel~A. Verheijen, David Busse,
  Claudia R{\"{o}}dl, J{\"{u}}rgen Furthm{\"{u}}ller, Friedhelm Bechstedt,
  Julian Stangl, Jonathan~J. Finley, Silvana Botti, Jos~E.M. Haverkort, and
  Erik~P.A.M. Bakkers.
\newblock {Direct-bandgap emission from hexagonal Ge and SiGe alloys}.
\newblock {\em Nature}, 580(7802):205--209, 2020.

\bibitem{norman2018perspective}
Justin~C Norman, Daehwan Jung, Yating Wan, and John~E Bowers.
\newblock Perspective: The future of quantum dot photonic integrated circuits.
\newblock {\em APL Photonics}, 3(3):030901, 2018.

\bibitem{chli2016}
S.~Chen, W.~Li, J.~Wu, Q.~Jiang, M.~Tang, S.~Shutts, S.N. Elliott,
  A.~Sobiesierski, A.J. Seeds, I.~Ross, P.M. Smowton, and H.~Liu.
\newblock {Electrically pumped continuous-wave III-V quantum dot lasers on
  silicon}.
\newblock {\em Nat. Photonics}, 10:307, 2016.

\bibitem{jung2017high}
Daehwan Jung, Justin Norman, MJ~Kennedy, Chen Shang, Bongki Shin, Yating Wan,
  Arthur~C Gossard, and John~E Bowers.
\newblock High efficiency low threshold current 1.3 $\mu$ m inas quantum dot
  lasers on on-axis (001) gap/si.
\newblock {\em Applied Physics Letters}, 111(12):122107, 2017.

\bibitem{tang2019integration}
Mingchu Tang, Jae-Seong Park, Zhechao Wang, Siming Chen, Pamela Jurczak, Alwyn
  Seeds, and Huiyun Liu.
\newblock Integration of iii-v lasers on si for si photonics.
\newblock {\em Progress in Quantum Electronics}, 66:1--18, 2019.

\bibitem{sost2016}
B.~Song, C.~Stagarescu, S.~Ristic, A.~Behfar, and J.~Klamkin.
\newblock {3D integrated hybrid silicon laser}.
\newblock {\em Opt. Express}, 24:10435, 2016.

\bibitem{crsa2017}
G.~Crosnier, D.~Sanchez, S.~Bouchoule, P.~Monnier, G.~Beaudoin, I.~Sagnes,
  R.~Raj, and F.~Raineri.
\newblock {Hybrid indium phosphide-on-silicon nanolaser diode}.
\newblock {\em Nat. Photon.}, 11:297, 2017.

\bibitem{huli2019}
Y.~Hu, D.~Liang, K.~Mikherjee, Y.~Li, C.~Zhang, G.~Kurcveil, X.~Huang, and R.G.
  Beausoleil.
\newblock {III/V-on-Si MQW lasers by using a novel photonic integration method
  of regrowth on a bonding template}.
\newblock {\em Light: Science \& Applications}, 8:93, 2019.

\bibitem{szha2019}
B.~Szelag, K.~Hassan, L.~Adelmini, E.~Ghegin, P.~Rodriguez, F.~Nemouchi,
  P.~Brianceau, E.~Vermande, A.~Schembri, D.~Carrara, P.~Cavali\'{e},
  F.~Franchin, M.-C. Roure, L.~Sanchez, C.~Jany, and S.~Olivier.
\newblock {Hybrid iii-V/Silicon Technology for Laser Integration on a 200-mm
  Fully CMOS-Compatible Silicon Photonics Platform}.
\newblock {\em IEEE J. Sel. Top. Quant. Electron.}, 25:8201210, 2019.

\bibitem{jito2020}
Y.~Jiao, J.~van~der Tol, V.~Pogoretskii, J.~van Engelen, A.A. Kashi,
  S.~Reniers, Y.~Wang, X.~Zhao, and W.~Yao et~al.
\newblock {Indium Phosphide Membrane Nanophotonic Integrated Circuits on
  Silicon}.
\newblock {\em Physica Status Solidi}, 217:1900606, 2020.

\bibitem{jubo2012}
J.~Justice, C.~Bower, M.~Meitl, M.B. Mooney, M.A. Gubbins, and B.~Corbett.
\newblock {Wafer-scale integration of group III-V lasers on silicon using
  transfer printing of epitaxial layers}.
\newblock {\em Nat. Photonics}, 6:610, 2012.

\bibitem{zhha2018}
J.~Zhang, B.~Haq, J.~O'Callaghan, A.~Gocalinska, E.~Pelucchi, A.J. Trindade,
  B.~Corbett, G.~Morthier, and G.~Roelkens.
\newblock {Transfer-printing-based integration of a III-V-on-silicon
  distributed feedback laser}.
\newblock {\em Opt. Express}, 26:8821, 2018.

\bibitem{zhang2019iii}
Jing Zhang, Grigorij Muliuk, Joan Juvert, Sulakshna Kumari, Jeroen Goyvaerts,
  Bahawal Haq, Camiel Op~de Beeck, Bart Kuyken, Geert Morthier, Dries
  Van~Thourhout, et~al.
\newblock Iii-v-on-si photonic integrated circuits realized using
  micro-transfer-printing.
\newblock {\em APL photonics}, 4(11):110803, 2019.

\bibitem{haxu2021}
Y.~Han, Y.~Xue, Z.~Yan, and K.M. Lau.
\newblock {Selectively Grown III-V Lasers for Integrated Si-Photonics}.
\newblock {\em J. Lightwave Technol.}, 39:940, 2021.

\bibitem{gurioli1991temperature}
M~Gurioli, A~Vinattieri, M~Colocci, C~Deparis, J~Massies, G~Neu, A~Bosacchi,
  and S~Franchi.
\newblock Temperature dependence of the radiative and nonradiative
  recombination time in gaas/al x ga 1- x as quantum-well structures.
\newblock {\em Physical Review B}, 44(7):3115, 1991.

\bibitem{dolores2017waveguide}
V~Dolores-Calzadilla, B~Romeira, F~Pagliano, S~Birindelli, A~Higuera-Rodriguez,
  PJ~Van~Veldhoven, MK~Smit, A~Fiore, and D~Heiss.
\newblock Waveguide-coupled nanopillar metal-cavity light-emitting diodes on
  silicon.
\newblock {\em Nature communications}, 8(1):1--8, 2017.

\bibitem{sa1958}
D.J. Sandiford.
\newblock {Temperature Dependence of Carrier Lifetime in Silicon}.
\newblock {\em Proc. Phys. Soc.}, 71:1002, 1958.

\bibitem{davies1989optical}
Gordon Davies.
\newblock The optical properties of luminescence centres in silicon.
\newblock {\em Physics reports}, 176(3-4):83--188, 1989.

\bibitem{suku2014}
H.~Sumikura, E.~Kuramochi, H.~Taniyama, and M.~Notomi.
\newblock Ultrafast spontaneous emission of copper-doped silicon enhanced by an
  optical nanocavity.
\newblock {\em Sci. Rep.}, 4:5040, 2014.

\bibitem{buckley2017all}
Sonia Buckley, Jeffrey Chiles, Adam~N McCaughan, Galan Moody, Kevin~L
  Silverman, Martin~J Stevens, Richard~P Mirin, Sae~Woo Nam, and Jeffrey~M
  Shainline.
\newblock All-silicon light-emitting diodes waveguide-integrated with
  superconducting single-photon detectors.
\newblock {\em Applied Physics Letters}, 111(14):141101, 2017.

\bibitem{bere2018}
C.~Beaufils, W.~Redjem, E.~Rousseau, V.~Jacques, A.Y. Kuznetsov, C.~Raynaud,
  C.~Voisin, A.~Benali, T.~Herzig, S.~Pezzagna, J.~Meijer, M.~Abbarchi, and
  G.~cassabois.
\newblock {Optical properties of an ensembe of G-centers in silicon}.
\newblock {\em Phys. Rev. B}, 97:035303, 2018.

\bibitem{chbe2018}
C.~Chartrand, L.~Bergeron, K.J. Morse, H.~Riemann, N.V. Abrosimov, P.~Becker,
  H.-J. Pohl, S.~Simmons, and M.L.W. Thewalt.
\newblock {Highly enriched $^28$Si reveals remarkable optical linewidths and
  fine structure for well-known damage centers}.
\newblock {\em Phys. Rev. B}, 98:195201, 2018.

\bibitem{hobe2020}
M.~Hollenback, Y.~Berenc\'{e}n, U.~Kentsch, M.~Helm, and G.V. Astakhov.
\newblock {Engineering telecom single-photon emitters in silicon for scalable
  quantum photonics}.
\newblock {\em Opt. Express}, 28:26111, 2020.

\bibitem{redu2020}
W.~Redjem, A.~Durand, T.~Herzig, A.~Benali, S.~Pezzagna, J.~Meijer, A.~Yu.
  Kuznetsov, H.S. Nguyen, S.~Cueff, J.-M. G\'{e}rard, I.~Robert-Philip, B.~Gil,
  D.~Caliste, P.~Pochet, M.~Abbarchi, V.~Jacques, A.~Dr\'{e}au, and
  G.~Cassabois.
\newblock {Single artificial atoms in silicon emitting at telecom wavelengths}.
\newblock {\em Nat. Electronics}, 3:738, 2020.

\bibitem{bech2020}
L.~Bergeron, C.~Chartrand, A.T.K. Kurkjian, K.J. Morse, H.~Riemann, N.V.
  Abrosimov, P.~Becker, H.-J. Pohl, M.L.W. Thewalt, and S.~Simmons.
\newblock {Silicon-Integrated Telecommunications Photon-Spin Interface}.
\newblock {\em PRX Quantum}, 1:020301, 2020.

\bibitem{romeira2018purcell}
Bruno Romeira and Andrea Fiore.
\newblock Purcell effect in the stimulated and spontaneous emission rates of
  nanoscale semiconductor lasers.
\newblock {\em IEEE Journal of Quantum Electronics}, 54(2):1--12, 2018.

\bibitem{bao2007point}
Jiming Bao, Malek Tabbal, Taegon Kim, Supakit Charnvanichborikarn, James~S
  Williams, Michael~J Aziz, and Federico Capasso.
\newblock Point defect engineered si sub-bandgap light-emitting diode.
\newblock {\em Optics Express}, 15(11):6727--6733, 2007.

\bibitem{buckley2020optimization}
Sonia~M Buckley, Alexander~N Tait, Galan Moody, Bryce Primavera, Stephen Olson,
  Joshua Herman, Kevin~L Silverman, Satyavolu~Papa Rao, Sae~Woo Nam, Richard~P
  Mirin, et~al.
\newblock Optimization of photoluminescence from w centers in a
  silicon-on-insulator.
\newblock {\em Optics express}, 28(11):16057--16072, 2020.

\bibitem{bowers2016recent}
John~E Bowers, Tin Komljenovic, Michael Davenport, Jared Hulme, Alan~Y Liu,
  Christos~T Santis, Alexander Spott, Sudharsanan Srinivasan, Eric~J Stanton,
  and Chong Zhang.
\newblock Recent advances in silicon photonic integrated circuits.
\newblock In {\em Next-Generation Optical Communication: Components,
  Sub-Systems, and Systems V}, volume 9774, page 977402. International Society
  for Optics and Photonics, 2016.

\bibitem{halbritter2014high}
Hubert Halbritter, Claus J{\"a}ger, Rolf Weber, Michael Schwind, and Frank
  M{\"o}llmer.
\newblock High-speed led driver for ns-pulse switching of high-current leds.
\newblock {\em IEEE Photonics Technology Letters}, 26(18):1871--1873, 2014.

\bibitem{kose2000}
C.~Koch and I~Segev.
\newblock {The Role of Single Neurons in Information Processing}.
\newblock {\em Nature Neuroscience}, 3:1171, 2000.

\bibitem{stsp2015}
G.J. Stuart and N.~Spruston.
\newblock Dendritic integration: 60 years of progress.
\newblock {\em Nature Neuroscience}, 18:1713, 2015.

\bibitem{haah2016}
J.~Hawkins and S.~Ahmad.
\newblock Why neurons have thousands of synapses, a theory of sequence memory
  in neocortex.
\newblock {\em Frontiers in Neural Circuits}, 10:23, 2016.

\bibitem{sava2017}
S.~Sardi, R.~Vardi, A.~Sheinin, A.~Goldental, and I.~Kanter.
\newblock New types of experiments reveal that a neuron functions as multiple
  independent threshold units.
\newblock {\em Scientific Reports}, 7:18036, 2017.

\bibitem{voma2007}
R.J. Vogelstein, U.~Mallik, J.T. Vogelstein, and G.~Cauwenberghs.
\newblock Dynamically reconfigurable silicon array of spiking neurons with
  conductance-based synapses.
\newblock {\em IEEE Trans. Neural Networks}, 18:253, 2007.

\bibitem{indiveri2011neuromorphic}
Giacomo Indiveri, Bernab{\'e} Linares-Barranco, Tara~Julia Hamilton, Andr{\'e}
  Van~Schaik, Ralph Etienne-Cummings, Tobi Delbruck, Shih-Chii Liu, Piotr
  Dudek, Philipp H{\"a}fliger, Sylvie Renaud, et~al.
\newblock Neuromorphic silicon neuron circuits.
\newblock {\em Frontiers in neuroscience}, 5:73, 2011.

\bibitem{pfgr2013}
T.~Pfeil, A.~Grubl, S.~Jeltsch, E.~M\"{u}ller, M.A. Metrovici, M.~Schmuker,
  D.~Br\"{u}derle, J.~Schemmel, and K.~Meier.
\newblock Six networks on a universal neuromorphic computing substrate.
\newblock {\em Frontiers in Neuroscience}, 7:1, 2013.

\bibitem{brne2013}
S.~Brink, S.~Nease, and P.~Hasler.
\newblock {Computing with networks of spiking neurons on a biophysically
  motivated floating-gate based neuromorphic integrated circuit}.
\newblock {\em Neural Networks}, 45:39, 2013.

\bibitem{bega2014}
B.V. Benjamin, P.~Gao, E.~McQuinn, S.~Choudhary, A.R. Chandresekaran, J.-M.
  Bussat, R.~Alvarez-Icaza, J.V. Arthur, P.A. Merolla, and K.~Boahen.
\newblock Neurogrid: A mixed-analog-digital multichip system for large-scale
  neural simulations.
\newblock {\em Proceedings of the IEEE}, 102:699, 2014.

\bibitem{abta2019}
K.~Abu-Hassan, J.D. Taylor, P.G. Morris, E.~Donati, Z.A. Bortolotto,
  G.~Indiveri, J.F.R. Paton, and A.~Nogaret.
\newblock Optimal solid state neurons.
\newblock {\em Nat. Comm.}, 10:5309, 2019.

\bibitem{crotty2010josephson}
Patrick Crotty, Dan Schult, and Ken Segall.
\newblock Josephson junction simulation of neurons.
\newblock {\em Physical Review E}, 82(1):011914, 2010.

\bibitem{toomey2019design}
Emily Toomey, Ken Segall, and Karl~K Berggren.
\newblock Design of a power efficient artificial neuron using superconducting
  nanowires.
\newblock {\em Frontiers in neuroscience}, 13:933, 2019.

\bibitem{mead1990neuromorphic}
Carver Mead.
\newblock Neuromorphic electronic systems.
\newblock {\em Proceedings of the IEEE}, 78(10):1629--1636, 1990.

\bibitem{rajendran2012specifications}
Bipin Rajendran, Yong Liu, Jae-sun Seo, Kailash Gopalakrishnan, Leland Chang,
  Daniel~J Friedman, and Mark~B Ritter.
\newblock Specifications of nanoscale devices and circuits for neuromorphic
  computational systems.
\newblock {\em IEEE Transactions on Electron Devices}, 60(1):246--253, 2012.

\bibitem{lide2015}
S.-C. Liu, T.~Delbruck, G.~Indiveri, A.~Whatley, and R.~Douglas, editors.
\newblock {\em Event-based neuromorphic systems}.
\newblock John Wiley and Sons, 2015.

\bibitem{sourikopoulos20174}
Ilias Sourikopoulos, Sara Hedayat, Christophe Loyez, Fran{\c{c}}ois Danneville,
  Virginie Hoel, Eric Mercier, and Alain Cappy.
\newblock A 4-fj/spike artificial neuron in 65 nm cmos technology.
\newblock {\em Frontiers in neuroscience}, 11:123, 2017.

\bibitem{indiveri2019importance}
Giacomo Indiveri and Yulia Sandamirskaya.
\newblock The importance of space and time for signal processing in
  neuromorphic agents: the challenge of developing low-power, autonomous agents
  that interact with the environment.
\newblock {\em IEEE Signal Processing Magazine}, 36(6):16--28, 2019.

\bibitem{schemmel2017accelerated}
Johannes Schemmel, Laura Kriener, Paul M{\"u}ller, and Karlheinz Meier.
\newblock An accelerated analog neuromorphic hardware system emulating nmda-and
  calcium-based non-linear dendrites.
\newblock In {\em 2017 International Joint Conference on Neural Networks
  (IJCNN)}, pages 2217--2226. IEEE, 2017.

\bibitem{be2007}
J.M. Beggs.
\newblock The criticality hypothesis: how local cortical networks might
  optimize information processing.
\newblock {\em Philosophical transactions of the Royal Society A}, 366:329,
  2007.

\bibitem{dalgaty2019hybrid}
Thomas Dalgaty, Melika Payvand, Filippo Moro, Denys~RB Ly, Florian
  Pebay-Peyroula, Jerome Casas, Giacomo Indiveri, and Elisa Vianello.
\newblock Hybrid neuromorphic circuits exploiting non-conventional properties
  of rram for massively parallel local plasticity mechanisms.
\newblock {\em APL Materials}, 7(8):081125, 2019.

\bibitem{chicca2014neuromorphic}
Elisabetta Chicca, Fabio Stefanini, Chiara Bartolozzi, and Giacomo Indiveri.
\newblock Neuromorphic electronic circuits for building autonomous cognitive
  systems.
\newblock {\em Proceedings of the IEEE}, 102(9):1367--1388, 2014.

\bibitem{hago1991}
Y.~Harada and E.~Goto.
\newblock Artificial neural network circuits with josephson devices.
\newblock {\em IEEE Trans. Magnetics}, 27:2863, 1991.

\bibitem{hiak1991}
M.~Hidaka and L.A. Akers.
\newblock An artificial neural cell implemented with superconducting circuits.
\newblock {\em Supercond. Sci. Technol.}, 4:654, 1991.

\bibitem{schneider2018tutorial}
Michael~L Schneider, Christine~A Donnelly, and Stephen~E Russek.
\newblock Tutorial: High-speed low-power neuromorphic systems based on magnetic
  josephson junctions.
\newblock {\em Journal of Applied Physics}, 124(16):161102, 2018.

\bibitem{schneider2017energy}
Michael~L Schneider, Christine~A Donnelly, Stephen~E Russek, Burm Baek,
  Matthew~R Pufall, Peter~F Hopkins, and William~H Rippard.
\newblock Energy-efficient single-flux-quantum based neuromorphic computing.
\newblock In {\em 2017 IEEE International Conference on Rebooting Computing
  (ICRC)}, pages 1--4. IEEE, 2017.

\bibitem{cocchi2017criticality}
Luca Cocchi, Leonardo~L Gollo, Andrew Zalesky, and Michael Breakspear.
\newblock Criticality in the brain: A synthesis of neurobiology, models and
  cognition.
\newblock {\em Progress in neurobiology}, 158:132--152, 2017.

\bibitem{schneider2020fan}
ML~Schneider and K~Segall.
\newblock Fan-out and fan-in properties of superconducting neuromorphic
  circuits.
\newblock {\em Journal of Applied Physics}, 128(21):214903, 2020.

\bibitem{tolpygo2016superconductor}
Sergey~K Tolpygo.
\newblock Superconductor digital electronics: Scalability and energy efficiency
  issues.
\newblock {\em Low Temperature Physics}, 42(5):361--379, 2016.

\bibitem{kisa2011}
D.E. Kirichenko, S.~Sarwana, and A.F. Kirichenko.
\newblock Zero static power dissipation biasing of rsfq circuits.
\newblock {\em IEEE Trans. Appl. Supercond.}, 21:776, 2011.

\bibitem{vrso1996}
C.~van Vreeswijk and H.~Sompolinsky.
\newblock Chaos in neuronal networks with balanced excitatory and inhibitory
  activity.
\newblock {\em Science}, 274:1724, 1996.

\bibitem{vora2005}
T.P. Vogels, K.~Rajan, and L.F. Abbott.
\newblock Neural network dynamics.
\newblock {\em Annu. Rev. Neurosci.}, 28:357, 2005.

\bibitem{zahoor2020resistive}
Furqan Zahoor, Tun~Zainal Azni~Zulkifli, and Farooq~Ahmad Khanday.
\newblock Resistive random access memory (rram): an overview of materials,
  switching mechanism, performance, multilevel cell (mlc) storage, modeling,
  and applications.
\newblock {\em Nanoscale research letters}, 15:1--26, 2020.

\bibitem{schneider2018ultralow}
Michael~L Schneider, Christine~A Donnelly, Stephen~E Russek, Burm Baek,
  Matthew~R Pufall, Peter~F Hopkins, Paul~D Dresselhaus, Samuel~P Benz, and
  William~H Rippard.
\newblock Ultralow power artificial synapses using nanotextured magnetic
  josephson junctions.
\newblock {\em Science advances}, 4(1):e1701329, 2018.

\bibitem{pfeil20124}
Thomas Pfeil, Tobias~C Potjans, Sven Schrader, Wiebke Potjans, Johannes
  Schemmel, Markus Diesmann, and Karlheinz Meier.
\newblock Is a 4-bit synaptic weight resolution enough?--constraints on
  enabling spike-timing dependent plasticity in neuromorphic hardware.
\newblock {\em Frontiers in neuroscience}, 6:90, 2012.

\bibitem{wang2018training}
Naigang Wang, Jungwook Choi, Daniel Brand, Chia-Yu Chen, and Kailash
  Gopalakrishnan.
\newblock Training deep neural networks with 8-bit floating point numbers.
\newblock {\em arXiv preprint arXiv:1812.08011}, 2018.

\bibitem{bartol2015nanoconnectomic}
Thomas~M Bartol~Jr, Cailey Bromer, Justin Kinney, Michael~A Chirillo,
  Jennifer~N Bourne, Kristen~M Harris, and Terrence~J Sejnowski.
\newblock Nanoconnectomic upper bound on the variability of synaptic
  plasticity.
\newblock {\em Elife}, 4:e10778, 2015.

\bibitem{fudr2005}
S.~Fusi, P.J. Drew, and L.F. Abbott.
\newblock Casdcade models of synaptically stored memories.
\newblock {\em Neuron}, 45:599, 2005.

\bibitem{fuab2007}
S.~Fusi and L.F. Abbott.
\newblock Limits on the memory storage capacity of bounded synapses.
\newblock {\em Nature Neuroscience}, 10:485, 2007.

\bibitem{upadhyay2019emerging}
Navnidhi~K Upadhyay, Hao Jiang, Zhongrui Wang, Shiva Asapu, Qiangfei Xia, and
  J~Joshua~Yang.
\newblock Emerging memory devices for neuromorphic computing.
\newblock {\em Advanced Materials Technologies}, 4(4):1800589, 2019.

\bibitem{diorio1998floating}
Chris Diorio, Paul Hasler, Bradley~A Minch, and Carver Mead.
\newblock Floating-gate mos synapse transistors.
\newblock In {\em Neuromorphic Systems Engineering}, pages 315--337. Springer,
  1998.

\bibitem{ramakrishnan2011floating}
Shubha Ramakrishnan, Paul~E Hasler, and Christal Gordon.
\newblock Floating gate synapses with spike-time-dependent plasticity.
\newblock {\em IEEE Transactions on Biomedical Circuits and Systems},
  5(3):244--252, 2011.

\bibitem{hasler2013finding}
Jennifer Hasler and Harry~Bo Marr.
\newblock Finding a roadmap to achieve large neuromorphic hardware systems.
\newblock {\em Frontiers in neuroscience}, 7:118, 2013.

\bibitem{stsn2008}
D.B. Strukov, G.S. Snider, D.R. Stewart, and R.S. Williams.
\newblock The missing memristor found.
\newblock {\em Nature}, 453:80, 2008.

\bibitem{yast2012}
J.J. Yang, D.B. Strukov, and D.R. Stewart.
\newblock Memristive devices for computing.
\newblock {\em Nat. Nanotech.}, 8:13, 2012.

\bibitem{ab2018}
I.~Abraham.
\newblock The case for rejecting the memristor as a fundamental circuit
  element.
\newblock {\em Nature}, 8:10972, 2018.

\bibitem{yin2019monolithically}
Shihui Yin, Yulhwa Kim, Xu~Han, Hugh Barnaby, Shimeng Yu, Yandong Luo, Wangxin
  He, Xiaoyu Sun, Jae-Joon Kim, and Jae-sun Seo.
\newblock Monolithically integrated rram-and cmos-based in-memory computing
  optimizations for efficient deep learning.
\newblock {\em IEEE Micro}, 39(6):54--63, 2019.

\bibitem{ielmini2018brain}
Daniele Ielmini.
\newblock Brain-inspired computing with resistive switching memory (rram):
  Devices, synapses and neural networks.
\newblock {\em Microelectronic Engineering}, 190:44--53, 2018.

\bibitem{ambrogio2016unsupervised}
Stefano Ambrogio, Nicola Ciocchini, Mario Laudato, Valerio Milo, Agostino
  Pirovano, Paolo Fantini, and Daniele Ielmini.
\newblock Unsupervised learning by spike timing dependent plasticity in phase
  change memory (pcm) synapses.
\newblock {\em Frontiers in neuroscience}, 10:56, 2016.

\bibitem{kim2019ferroelectric}
Min-Kyu Kim and Jang-Sik Lee.
\newblock Ferroelectric analog synaptic transistors.
\newblock {\em Nano letters}, 19(3):2044--2050, 2019.

\bibitem{kim2018recent}
Sun~Gil Kim, Ji~Su Han, Hyojung Kim, Soo~Young Kim, and Ho~Won Jang.
\newblock Recent advances in memristive materials for artificial synapses.
\newblock {\em Advanced Materials Technologies}, 3(12):1800457, 2018.

\bibitem{zhang2020brain}
Yang Zhang, Zhongrui Wang, Jiadi Zhu, Yuchao Yang, Mingyi Rao, Wenhao Song,
  Ye~Zhuo, Xumeng Zhang, Menglin Cui, Linlin Shen, et~al.
\newblock Brain-inspired computing with memristors: Challenges in devices,
  circuits, and systems.
\newblock {\em Applied Physics Reviews}, 7(1):011308, 2020.

\bibitem{vatu1998}
T.~Van Duzer and C.W. Turner.
\newblock {\em Principles of superconductive devices and circuits}.
\newblock Prentice Hall, USA, second edition, 1998.

\bibitem{ka1999}
Alan~M. Kadin.
\newblock {\em Introduction to superconducting circuits}.
\newblock John Wiley and Sons, USA, first edition, 1999.

\bibitem{sh2018}
J.M. Shainline, S.M. Buckley, A.N. McCaughan, J.~Chiles, A.~Jafari-Salim, R.P.
  Mirin, and S.W. Nam.
\newblock Circuit designs for superconducting optoelectronic loop neurons.
\newblock {\em J. Appl. Phys.}, page 152130, 2018.

\bibitem{tobo2018}
S.K. Tolpygo, V.~Bolkhovsky, D.E. Oates, R.~Rastogi, S.~Zarr, A.L. Day, T.J.
  Weir, A.~Wynn, and L.M. Johnson.
\newblock {Superconductor Electronics Fabrication Process with MoN$_x$ Kinetic
  Inductors and Self-Shunted Josephson Junctions}.
\newblock {\em IEEE Transactions on Applied Superconductivity}, 28(4):1100212,
  2018.

\bibitem{brsc1998}
V.~Braitenberg and A.~Schuz.
\newblock {\em Cortex: statistics and geometry of neuronal connectivity}.
\newblock Springer, Berlin, Germany, 1998.

\bibitem{bu2006}
G.~Buzs\'{a}ki.
\newblock {\em Rhythms of the Brain}.
\newblock Oxford University Press, 2006.

\bibitem{busp2012}
E.~Bullmore and O.~Sporns.
\newblock The economy of brain network organization.
\newblock {\em Nature Reviews Neuroscience}, 13:336, 2012.

\bibitem{frfr2004}
A.~Fronczak, P.~Fronczak, and J.A. Holyst.
\newblock Average path length in random networks.
\newblock {\em Phys. Rev. E}, 70:056110, 2004.

\bibitem{si1962}
H.A. Simon.
\newblock The {A}rchitecture of {C}omplexity.
\newblock {\em Proc. Amer. Phil. Soc.}, 106:467, 1962.

\bibitem{mela2010}
D.~Meunier, R.~Lambiotte, and E.T. Bullmore.
\newblock {Modular and hierarchically modular organization of brain networks}.
\newblock {\em Frontiers in Neuroscience}, 4(200):1, 2010.

\bibitem{mo1997}
V.B. Mountcastle.
\newblock The columnar organization of the neocortex.
\newblock {\em Brain}, 120:701, 1997.

\bibitem{bosp2015}
M.~Bota, O.~Sporns, and L.W. Swanson.
\newblock Architecture of the cerebral cortical association connectome
  underlying cognition.
\newblock {\em PNAS}, page E2093, 2015.

\bibitem{beba2017}
R.F. Betzel and D.S. Bassett.
\newblock Multi-scale brain networks.
\newblock {\em NeuroImage}, 160:73, 2017.

\bibitem{busp2009}
E.~Bullmore and O.~Sporns.
\newblock Complex brain networks: graph theoretical analysis of structural and
  functional systems.
\newblock {\em Nature Reviews Neuroscience}, 10:186, 2009.

\bibitem{frbu2016}
K.~Friston and G.~Buzs\'{a}ki.
\newblock {The Functional Anatomy of Time: What and When in the Brain}.
\newblock {\em Trends in Cognitive Sciences}, 20(7):500, 2016.

\bibitem{ke1982}
R.W. Keyes.
\newblock The wire-limited logic chip.
\newblock {\em IEEE J. Sol.-Sta. Circuits}, SC-17:1232, 1982.

\bibitem{schemmel2010wafer}
Johannes Schemmel, Daniel Br{\"u}derle, Andreas Gr{\"u}bl, Matthias Hock,
  Karlheinz Meier, and Sebastian Millner.
\newblock A wafer-scale neuromorphic hardware system for large-scale neural
  modeling.
\newblock In {\em 2010 IEEE International Symposium on Circuits and Systems
  (ISCAS)}, pages 1947--1950. IEEE, 2010.

\bibitem{shpa2015}
K.~Shang, S.~Pathak, B.~Guan, G.~Liu, and S.J.B. Yoo.
\newblock {Low-loss compact multilayer silicon nitride platform for 3D photonic
  integrated circuits}.
\newblock {\em Opt. Express}, 23:21334, 2015.

\bibitem{sahu2015}
W.D. Sacher, Y.~Huang, G.-Q. Lo, and J.K.S. Poon.
\newblock Multilayer silicon nitride-on-silicon integrated photonic platforms
  and devices.
\newblock {\em J. Lightwave Tech.}, 33:901, 2015.

\bibitem{chbu2017}
J.~Chiles, S.~Buckley, N.~Nader, S.W. Nam, R.P. Mirin, and J.M. Shainline.
\newblock Multi-planar amorphous silicon photonics with compact interplanar
  couplers, cross talk mitigation, and low crossing loss.
\newblock {\em APL Photonics}, 2:116101, 2017.

\bibitem{zhli2018}
Y.~Zhang, Y.~Ling, Y.~Zhang, K.~Shang, and S.J.B. Yoo.
\newblock High-density wafer-scale 3-d silicon-photonic integrated circuits.
\newblock {\em IEEE J. Sel. Topics Quantum Electron.}, 24:8200510, 2018.

\bibitem{ro1983}
A.L. Rosenberg.
\newblock {Three-Dimensional VLSI: A Case Study}.
\newblock {\em J. Assoc. Computing Machinery}, 30(3):397, 1983.

\bibitem{knan2008}
J~U Knickerbocker, P~S Andry, B~Dang, R~R Horton, M~J Interrante, C~S Patel,
  R~J Polastre, K~Sakuma, R~Sirdeshmukh, E~J Sprogis, A~M Stephens, A~W Topol,
  C~K Tsang, B~C Webb, and S~L Wright.
\newblock {Three- dimensional silicon integration}.
\newblock {\em IBM J. Res. \& Dev.}, 52(6):553--569, 2008.

\bibitem{saan2008}
K.~Sakuma, P.S. Andry, C.K. Tsang, S.L. Wright, B.~Dang, C.S. Patel, B.C. Webb,
  J.~Maria, E.J. Sprogis, S.K. Kang, R.J. Polastre, R.R. Horton, and J.U.
  Knickerbocker.
\newblock {3D chip-stacking technology with through-silicon vias and low-volume
  lead-free interconnections}.
\newblock {\em IBM J. Res. \& Dev.}, 52(6):611, 2008.

\bibitem{viba2011}
M.~Vinet et~al.
\newblock {3D monolithic integration: Technological challenges and electrical
  results}.
\newblock {\em Microelectronic Engineering}, 88:331, 2011.

\bibitem{zhxi2015}
J.~Zhao, Y.~Xie, and Q.~Zou.
\newblock {Overview of 3-D Architecture Design Opportunities and Techniques}.
\newblock {\em IEEE Design \& Test}, page~60, 2015.

\bibitem{li2013}
S.K. Lim.
\newblock {\em Design for High Performance, Low Power, and Reliable 3D
  Integrated Circuits}.
\newblock Springer, 2013.

\bibitem{elfe2016}
{\em {3D Stacked Chips}}.
\newblock Springer, 2016.

\bibitem{lish2017}
M.~Li, J.~Shi, M.~Rahman, S.~Khasanvis, S.~Bhat, and C.A. Moritz.
\newblock {Skybridge-3D-CMOS: A Fine-Grained 3D CMOS Integrated Circuit
  Technology}.
\newblock {\em IEEE Trans. Nanotech.}, 16(4):639, 2017.

\bibitem{rupa2019}
Arianna Rubino, Melika Payvand, and Giacomo Indiveri.
\newblock {2019 26th IEEE International Conference on Electronics, Circuits and
  Systems, ICECS 2019}.
\newblock {\em 2019 26th IEEE International Conference on Electronics, Circuits
  and Systems, ICECS 2019}, pages 458--461, 2019.

\bibitem{tobo2019}
S.K. Tolpygo, V.~Bolkhovsky, R.~Rastogi, S.~Zarr, A.L. Day, E.~Golden, T.J.
  Weir, A.~Wynn, and L.M. Johnson.
\newblock {Planarized Fabrication Process With Two Layers of SIS Josephson
  Junctions and Integration of SIS and SFS $\pi$-Junctions}.
\newblock {\em IEEE Transactions on Applied Superconductivity}, 29(5):1101208,
  2019.

\bibitem{anna2017}
T.~Ando, S.~Nagasawa, N.~Takeuchi, N.~Tsuji, F.~China, M.~Hidaka, Y.~Yamanashi,
  and N.~Yoshikawa.
\newblock {Three-dimensional adiabatic quantum-flux-parametron fabricated using
  a double-active-layered niobium process}.
\newblock {\em Supercond. Sci. Technol.}, 30:075003, 2017.

\bibitem{shainline2020optoelectronic}
Jeffrey~M Shainline.
\newblock Optoelectronic intelligence.
\newblock {\em arXiv preprint arXiv:2010.08690}, 2020.

\bibitem{mo1978}
V.B. Mountcastle.
\newblock {\em An {O}rganizing {P}rinciple for {C}erebral {F}unction: {T}he
  {U}nit {M}odule and the {D}istributed {S}ystem}.
\newblock The MIT Press, 1978.

\bibitem{alekseev2015basics}
A~Alekseev.
\newblock Basics of low-temperature refrigeration.
\newblock {\em arXiv preprint arXiv:1501.07392}, 2015.

\bibitem{holmes2013energy}
D~Scott Holmes, Andrew~L Ripple, and Marc~A Manheimer.
\newblock Energy-efficient superconducting computing—power budgets and
  requirements.
\newblock {\em IEEE Transactions on Applied Superconductivity},
  23(3):1701610--1701610, 2013.

\bibitem{zhirnov2003limits}
Victor~V Zhirnov, Ralph~K Cavin, James~A Hutchby, and George~I Bourianoff.
\newblock Limits to binary logic switch scaling-a gedanken model.
\newblock {\em Proceedings of the IEEE}, 91(11):1934--1939, 2003.

\bibitem{linares2003design}
Bernab{\'e} Linares-Barranco and Teresa Serrano-Gotarredona.
\newblock On the design and characterization of femtoampere current-mode
  circuits.
\newblock {\em ieee journal of solid-state circuits}, 38(8):1353--1363, 2003.

\bibitem{wu2009metal}
Yung-Hsien Wu, Bo-Yu Chen, Lun-Lun Chen, Jia-Rong Wu, and Min-Lin Wu.
\newblock Metal-insulator-metal capacitor with high capacitance density and low
  leakage current using zrtio 4 film.
\newblock {\em Applied Physics Letters}, 95(11):113502, 2009.

\bibitem{matula1979electrical}
Richard~Allen Matula.
\newblock Electrical resistivity of copper, gold, palladium, and silver.
\newblock {\em Journal of Physical and Chemical Reference Data},
  8(4):1147--1298, 1979.

\bibitem{clbr2006}
J.~Clarke and A.I. Braginski, editors.
\newblock {\em The SQUID handbook}.
\newblock Wiley-VCH.

\bibitem{jaycox1981planar}
J~Jaycox and M~Ketchen.
\newblock Planar coupling scheme for ultra low noise dc squids.
\newblock {\em IEEE Transactions on Magnetics}, 17(1):400--403, 1981.

\end{thebibliography}
\end{document}